\documentclass{article}
\usepackage{amssymb}
\usepackage{cite,multirow} 
\usepackage[dvips]{graphicx,epsfig}
\usepackage{graphics}
\def\1ad{\mbox{\normalsize $^1$}}
\def\2ad{\mbox{\normalsize $^2$}}
\def\3ad{\mbox{\normalsize $^3$}}
\def\4ad{\mbox{\normalsize $^4$}}
\def\5ad{\mbox{\normalsize $^5$}}
\def\6ad{\mbox{\normalsize $^6$}}
\def\7ad{\mbox{\normalsize $^7$}}
\def\8ad{\mbox{\normalsize $^8$}}

\makeatletter
\@addtoreset{equation}{section}
\makeatother

\newcommand{\ignora}[1]{}
\def\beq{\begin{equation}}                     %
\def\eeq{\end{equation}}                       %
\def\bea{\begin{eqnarray}}                     
\def\eea{\end{eqnarray}}                       
\def\0 {\nonumber}

\begin{document}

\setcounter{page}{0}
\begin{titlepage}
\titlepage
\rightline{hep-th/0611229}
\rightline{Bicocca-FT-06-18}
\rightline{LPTENS-06/54}
\rightline{SISSA 65/2006/EP}
\vskip 3cm
\centerline{{ \bf \Large Counting BPS Baryonic Operators in CFTs}}
\vskip 0.4cm
\centerline{{ \bf \Large with Sasaki-Einstein duals}}
\vskip 1cm
\centerline{{\bf Agostino Butti  $^{a}$, Davide Forcella  $^{b}$, Alberto Zaffaroni  $^{c}$}}
\vskip 1truecm
\begin{center}
\em 
 $^{a}$ Laboratoire de Physique Th\'eorique de l'\'Ecole Normale Sup\'erieure\\
24, rue Lhomond, 75321 Paris Cedex 05, France\\
\vskip 0.5cm
$^{b}$ International School for Advanced Studies (SISSA / ISAS)\\ 
via Beirut 2, I-34014, Trieste, Italy\\
\vskip 0.5cm
$^{c}$ Universit\`{a} di Milano-Bicocca and INFN, sezione di  Milano-Bicocca\\ 
P.zza della Scienza, 3; I-20126 Milano, Italy\\
\vskip .4cm

\vskip 2.5cm
\end{center}
\begin{abstract}
We study supersymmetric D3 brane configurations wrapping internal cycles
of type II backgrounds $AdS_5\times H$ for a generic Sasaki-Einstein
manifold H. These configurations correspond to BPS baryonic operators
in the dual quiver gauge theory. In each sector with given baryonic charge,
we write explicit partition functions counting all the BPS operators according to their flavor and R-charge. We also show how to extract geometrical information about H from the partition functions; in particular, we give general
formulae for computing volumes of three cycles in H.
\vskip1cm

\end{abstract}
\vskip 0.5\baselineskip

\vfill
 \hrule width 5.cm
\vskip 2.mm
{\small
\noindent agostino.butti@lpt.ens.fr\\
forcella@sissa.it\\
alberto.zaffaroni@mib.infn.it}
\begin{flushleft}
\end{flushleft}
\end{titlepage}
\large
\tableofcontents
\section{Introduction}

The study of $BPS$ states in quantum field theory and in string theory is clearly a very important topic. 
These states are generically protected against quantum corrections and contain information regarding the 
strong coupling behaviour of supersymmetric field theories and superstring theories. 
In the past years they were especially important in the study of strong weak dualities, like the AdS/CFT conjecture \cite{Maldacena:1997re} 
which gives a connection between the $BPS$ operators in conformal field  theories and $BPS$ states in string theory. 

In this paper we discuss the set of one half $BPS$ states in string theory realized as $D3$ branes wrapped on 
(generically non trivial) three cycles in the supergravity background $AdS_5 \times H$, where $H$ is a Sasaki-Einstein manifold \cite{kw,horizon}.
These states are holographically dual to baryonic $BPS$ operators in $\mathcal{N}=1$ four dimensional 
$CFT$s \cite{Gubser:1998fp}, which are quiver gauge theories. 

Recently, there has been renewed interest in generalizing
the $AdS/CFT$  correspondence to generic Sasaki-Einstein manifolds $H$. 
This interest has been initially motivated by the discovery of new infinite classes of non compact $CY$ metrics \cite{gauntlett,CLPP,MSL,MS} and 
the construction of their dual  $\mathcal{N}=1$ supersymmetric $CFT$ \cite{benvenuti,kru2,noi,tomorrow} 
\footnote{See \cite{force}, and references therein, for an overview of analogous
results for non-conformal fields theories.}. As a result of this line of
investigation, we now have a well defined correspondence between toric CY and  
dual quiver gauge theories 
\cite{kru2,tomorrow,bertolini,dimers,hananyX,MSY,kru,aZequiv,proc,rhombi,mirror}. The non toric case is still less understood: 
there exist studies on generalized conifolds \cite{Gubser,Esperanza}, del Pezzo series \cite{Hanany:2001py,wijnholt,Franco:2004rt}, 
and more recently there was a proposal to construct new non toric examples \cite{Butti:2006nk}.


There has been some parallel interest in counting $BPS$ states in the $CFT$s dual to CY singularities \cite{Romelsberger:2005eg,Kinney:2005ej,Biswas:2006tj,Mandal:2006tk,Benvenuti:2006qr,Martelli:2006vh}. The partition function counting mesonic $BPS$ gauge invariant operators according to their flavor quantum  numbers contains a lot of information regarding the geometry of the $CY$ 
\cite{Benvenuti:2006qr,MSY2}, including the algebraic equations of the singularity. Quite interestingly, it also provides a formula for the volume of $H$
\cite{MSY2}. This geometrical information has a direct counterpart in field theory, since, according to  the $AdS/CFT$ correspondence, the 
volume of the total space and of the three cycles are duals to the central 
charge and the $R$ charges of the baryonic operators respectively \cite{Gubser:1998fp,Gubser:1998vd}.

The existing countings focus on the mesonic gauge invariant sector of the $CFT$. Geometrically this corresponds to consider giant graviton configurations \cite{McGreevy:2000cw} corresponding to $BPS$ $D3$ branes wrapped on trivial three cycles in $H$. 
In this paper we push this investigation further and we analyze the baryonic
$BPS$ operators, corresponding to $D3$ branes wrapped on non trivial three cycles inside $H$. 
We succeed in counting $BPS$ states charged under the baryonic charges of the field theory and we write explicit partition functions at fixed
baryonic charge. We investigate in details their geometrical properties.
In particular we will show how to extract 
from the baryonic partition functions a formula for 
the volume of the three cycles inside $H$. 
We will mostly concentrate on the toric case but our procedure seems adaptable to the non toric case as well.

The paper is organized as follows.
In Section 2 we review some basic elements of toric geometry. In Section 3
we formulate the general problem of describing and quantizing the $BPS$ 
$D3$ brane configurations.  
We will use homomorphic surfaces to parameterize the supersymmetric $BPS$ configurations of $D3$ brane wrapped in $H$, following 
results in \cite{Mikhailov:2000ya,Beasley:2002xv} \footnote{See \cite{Canoura:2005uz,Canoura:2006es} for some recent developments 
in wrapping branes on non trivial three cycles inside toric singularities.}. 
In the case where $X$ is a toric variety we have globally defined homogeneous 
coordinates $x_i$ which are charged under the baryonic charges of the theory and which we can use to parametrize these surfaces. 
We will quantize configurations of D3 branes wrapped on these surfaces and we will find the Hilbert space of $BPS$ states 
using a prescription found by Beasley \cite{Beasley:2002xv}. The complete $BPS$ Hilbert space factorizes in sectors with 
definite baryonic charges. Using toric geometry tools, 
we can assign to each sector a convex polyhedron $P$. 
The BPS operators in a given sector are in one-to-one correspondence with 
symmetrized products of N (number of colors) integer points in $P$.
In Section 4 we discuss the assignment of charges and we set the general counting problem. In Section 5 we make some comparison with the field theory side. 
In Section 6, we will write a partition function $Z_D$ counting the integer points in $P_D$ and a partition function $Z_{D,N}$ counting 
the integer points in the symmetric product of $P_D$. 
From $Z_D$, taking a suitable limit, 
we will be able to compute the volume of the three cycles in $H$, as described in Section 7. 
Although we mainly focus on the toric case we propose a general formula for the computation of the volume of the three cycles 
valid for every type of conical $CY$ singularity.
   
From the knowledge of $Z_{D,N}$ we can 
reconstruct the complete partition function for the chiral ring of quiver gauge theories. This is a quite hard problem in
field theory, since we need to count gauge invariant 
operators modulo F-term relations and to take into account the finite number of colors $N$ which induces relations among traces and determinants. The 
geometrical computation of $Z_{D,N}$ should allow to by-pass these problems. 
In this paper we will mainly focus on the geometrical properties of the partition functions $Z_D$, although some preliminary 
comparison with the dual gauge theory  is made in Section 5. In forthcoming
papers, we will show how to compute the complete partition function for selected examples and how to compare with field theory expectations \cite{ADAZ}.

\section{A short review of toric geometry}\label{toric} 

In this section we summarize some basic topics of toric geometry; in particular we review divisors and line bundles on toric 
varieties that will be very useful for the complete understanding of the paper. Very useful references on
toric geometry are \cite{fulton,cox}.

\begin{figure}[h!!!]
\begin{center}
\includegraphics[scale=0.6]{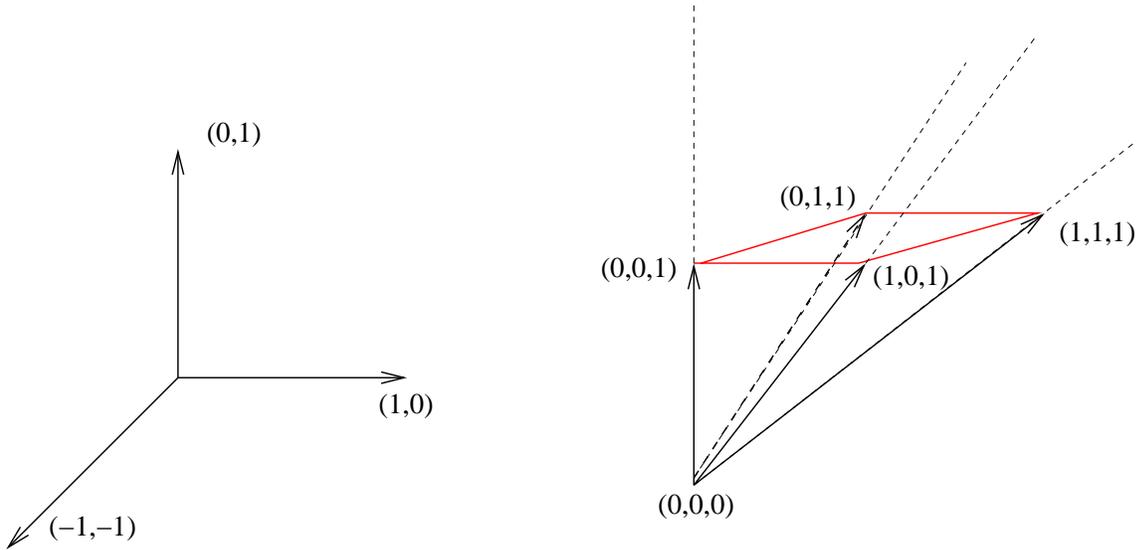} 
\caption{On the left: the fan for $\mathbb{P}^2$ with three maximal cones
of dimension two
which fill completely $\mathbb{R}^2$; there are three one dimensional cones
in $\Sigma(1)$ with generators $\{(1,0),(0,1),(-1,-1)\}$. On the right: the
fan for the conifold with a single maximal cone of dimension three; there are
four one dimensional cones  in $\Sigma(1)$ with generators $\{ (0,0,1),(1,0,1),(1,1,1),(0,1,1)\}$.}\label{fan}
\end{center}
\end{figure}

A toric variety $V_{\Sigma}$ is defined by a fan $\Sigma$: a collection of strongly  convex rational polyhedral cones in the 
real vector space $N_\mathbb{R} = N \otimes _ {\mathbb{Z}} \mathbb{R}$ ($N$ is an $n$ dimensional lattice $N \simeq \mathbb{Z}^n$). 
Some examples are presented in Figure \ref{fan}.

We define the variety $V_{\Sigma}$ as a symplectic quotient \cite{fulton,cox}. Consider the one dimensional cones of 
$\Sigma $ and a minimal integer generator $n_i$ of each of them. Call the set of one dimensional cones $\Sigma (1) $. 
Assign a ``homogeneous coordinate'' $x_i$ to each $n_i \in \Sigma (1) $.
If $d=$ dim$\Sigma (1) $, $x_i $ span $\mathbb{C}^d$. 
Consider the group
\begin{equation}
G=\{(\mu_1,...,\mu_d\}\in (\mathbb{C}^*)^d | \prod_{i=1}^d \mu_i^{<m,n_i>} =1\, ,\,\,\,\, m\in \mathbb{Z}^3\}\, ,
\label{Ggroup}\end{equation}
which acts on $x_i$ as
$$(x_1,...,x_d) \rightarrow (\mu_1 x_1,...,\mu_d x_d)\, .$$
 $G$ is isomorphic, in general, to $(\mathbb{C}^*)^{d-n}$ times a discrete group. The continuous part $(\mathbb{C}^*)^{d-n}$ 
can be described as follows.
Since $d \geqslant n $ the $n_i $ are not linearly independent. They determine $d-n$ linear  relations:
\begin{equation}\label{symplq}
\sum_{i=1}^{d} Q^{(a)}_i n_i = 0  
\end{equation}
with $a=1,...,d-n$ and $Q_i^{(a)}$ generate a $(\mathbb{C}^*)^{d-n}$ action on $\mathbb{C}^d$:
\begin{equation}(x_1,...,x_d) \rightarrow (\mu^{Q_1^{(a)}}x_1,...,\mu^{Q_d^{(a)}}x_d)\label{resc}\end{equation}
where $\mu \in \mathbb{C}^*$. 

For each maximal cone $\sigma \in \Sigma $ define the function $f_{\sigma}= \prod_{n_i \notin \sigma} x_i$ and the locus 
$S$ as the intersection of all the hypersurfaces $f_{\sigma}=0$. Then the toric variety is defined as: 
$$V_{\Sigma}= ( \mathbb{C}^d - S )/ G$$
There is a 
residual $(\mathbb{C}^*)^{n}$ complex torus action acting on $V_{\Sigma}$, from
which the name {\it toric variety}. In the following, we will denote with $T^n\equiv U(1)^n$ the real torus contained in $(\mathbb{C}^*)^{n}$.

In all the examples in this paper $G=(\mathbb{C}^*)^{d-n}$ and the previous quotient is interpreted as a symplectic reduction. 
The case where $G$ contains a discrete part includes further orbifold quotients. These cases can be handled similarly to the ones 
discussed in the main text. 

Using these rules to construct the toric variety, it is easy to recover
 the usual representation for $\mathbb{P}^n $:
$$\mathbb{P}^n = (\mathbb{C}^{n+1} - \{ 0 \})/ \{ x \sim \mu x \}$$
where the the minimal integer generators of $\Sigma (1)$ are $n_i = \{ e_1 ,..., e_n, -\sum_{k=1}^n e_k \}$, $d=n+1$ and 
$Q = (1,...,1)$ (see Figure \ref{fan} for the case $n=2$).

In this paper we will be interested in affine toric varieties, where the fan is a single cone $\Sigma$ = $\sigma$. 
In this case $S$ is always the null set. It is easy, for example, to find the symplectic quotient representation of the conifold:
$$C(T^{1,1}) = \mathbb{C}^4 / (1,-1,1,-1)$$ 
where $d=4$, $n=3$, $n_1=(0,0,1)$, $n_2=(1,0,1)$, $n_3=(1,1,1)$, $n_4=(0,1,1)$ and we have written $(1,-1,1,-1)$ for 
the action of $\mathbb{C}^*$ with charges $Q= (1,-1,1,-1)$. 

This type of description of a toric variety is the easiest one to study divisors and line bundles. Each $n_i \in \Sigma (1)$ 
determines a $T$-invariant divisor $D_i$ corresponding to the zero locus $\{ x_i = 0 \}$ in $V_{\Sigma}$. $T$-invariant means 
that $D_i$ is mapped to itself by the torus action $(\mathbb{C}^*)^{n}$ (for simplicity we will call them simply divisors from now on). 
The $d$ divisors $D_i$ are not independent but satisfy the $n$ basic equivalence relations:
\begin{equation}\label{eqrel}
\sum _{i=1}^d < e_k , n_i > D_i = 0
\end{equation}
where $e_k$ with $k= 1,...,n$ is the orthonormal basis of the dual lattice $M \sim \mathbb{Z}^n $ with the natural paring: 
for $n\in N$, $m\in M$ $<n,m> = \sum _{i=1}^n n_i m_i$ $\in \mathbb{Z}$.
Given the basic divisors $D_i$ the generic divisor $D$ is given by the formal
sum $D = \sum_{i=1}^d c_i D_i$ with $c_i \in \mathbb{Z}$. Every divisor $D$
determines a line bundle $\mathcal{O}(D)$ \footnote{The generic divisor $D$ on an
affine cone is a Weil divisor and not a Cartier divisor \cite{fulton}; for this reason the map between divisors 
and line bundles is more subtle, but it can be easily generalized using the homogeneous coordinate ring of the toric 
variety $V_{\Sigma}$ \cite{Cox:1993fz} in a way that we will explain. 
With an abuse of language, we will continue to call the sheaf $\mathcal{O}(D)$ the line bundle associated with the divisor $D$.}.

There exists a simple recipe to find the holomorphic sections of
 the line bundle $\mathcal{O}(D)$.
Given the $c_i$, the global sections of $\mathcal{O}(D)$ can be determined by looking at the polytope (a convex rational polyhedron in $M_{\mathbb{R}}$):
\begin{equation}\label{poly}
P_D = \{ u \in M_{\mathbb{R}}| <u,n_i >\hbox{  }\geq \hbox{  } - c_i \hbox{  }, \hbox{  }\forall i \in \Sigma(1) \}
\end{equation} 
where $ M_{\mathbb{R}}= M \otimes _ {\mathbb{Z}} \mathbb{R}$.
Using the homogeneous coordinate $x_i$ it is easy to associate a section $\chi^m$ to every point $m$ in $P_D$:
\begin{equation}\label{sect}
\chi ^m = \prod _{i=1}^d x_i ^{< m , n_i > + c_i} .
\end{equation} 
Notice that the exponent is equal or bigger than zero.
Hence the global sections of the line bundle $\mathcal{O}(D)$ over $V_{\Sigma}$ are:
\begin{equation}\label{sectgen}
H^0 ( V_{\Sigma}, \mathcal{O}_{V_{\Sigma}}(D))= \bigoplus _{m \in P_D \cap M } \mathbb{C} \cdot \chi ^m 
\end{equation} 
At this point it is important to make the following observation: all monomials $\chi ^m$ have the same charges under the $(\mathbb{C}^*)^{d-n}$ 
described at the beginning of this Section (in the following these charges will be identified with the baryonic charges of the dual gauge theory). 
Indeed, under the $(\mathbb{C}^*)^{d-n}$ action we have:
\begin{equation}\label{baryonicaction}
\chi ^m \rightarrow \prod _{i=1}^d(\mu ^{<m, Q_i^{(a)}n_i>+Q_i^{(a)}c_i})x_i ^{< m , n_i > + c_i} = \mu ^{\sum_{i=1}^d Q_i^{(a)} c_i} \chi ^m 
\end{equation} 
where we have used equation (\ref{symplq}). Similarly, all the sections have 
the same charge under the discrete part of the group $G$. 
This fact has an important consequence. The generic polynomial 
$$f = \sum a_m \chi^m \in H^0 ( V_{\Sigma}, \mathcal{O}_{V_{\Sigma}}(D))$$
is not a {\it function} on $V_{\Sigma}$, since it is not invariant under the $(\mathbb{C}^*)^{d-n}$ action (and under possible discrete orbifold actions).
However, it makes perfectly sense to consider the zero locus of $f$. Since all monomials in $f$ have the same charge under  $(\mathbb{C}^*)^{d-n}$, 
the equation $f=0$ is well defined on $V_{\Sigma}$ and defines a divisor \footnote{In this way, we can set a map between linearly equivalent divisors 
and sections of the sheaf $\mathcal{O}_{V_{\Sigma}}(D)$ generalizing the usual
map in the case of standard line bundles.}.

\subsection{A simple Example}
After this general discussion, let us discuss an example to clarify 
the previous definitions. 

Consider the toric variety $\mathbb{P}^2$. 
The fan $\Sigma $ for $\mathbb{P}^2$ is generated by:
\begin{equation}
n_1 = e_1 \hbox{   } n_2 = e_2 \hbox{   } n_3 = - e_1 - e_2 
\end{equation}
The three basic divisors $D_i$ correspond to $\{ x_1 = 0 \}  $, $\{ x_2 = 0 \}  $, $\{ x_3 = 0 \}  $, and they satisfy the following relations 
(see equation (\ref{eqrel})):
\begin{eqnarray}
D_1 - D_3 = 0 \nonumber \\
D_2 - D_3 = 0 \nonumber
\end{eqnarray}
and hence $D_1 \sim D_2 \sim D_3\sim D$. All line bundles on $\mathbb{P}^2$
are then of the form ${\cal O}(n D)$ with an integer $n$, and are usually
denoted as ${\cal O}(n)\rightarrow \mathbb{P}^2$. It is well known that  
the space of global holomorphic sections of ${\cal O}(n)\rightarrow \mathbb{P}^2$ is
given by the homogeneous polynomial of degree $n$ for $n\ge 0$, while
it is empty for negative $n$. We can verify this statement using the 
general construction with polytopes.

\begin{figure}[h!!!]
\begin{center}
\includegraphics[scale=0.6]{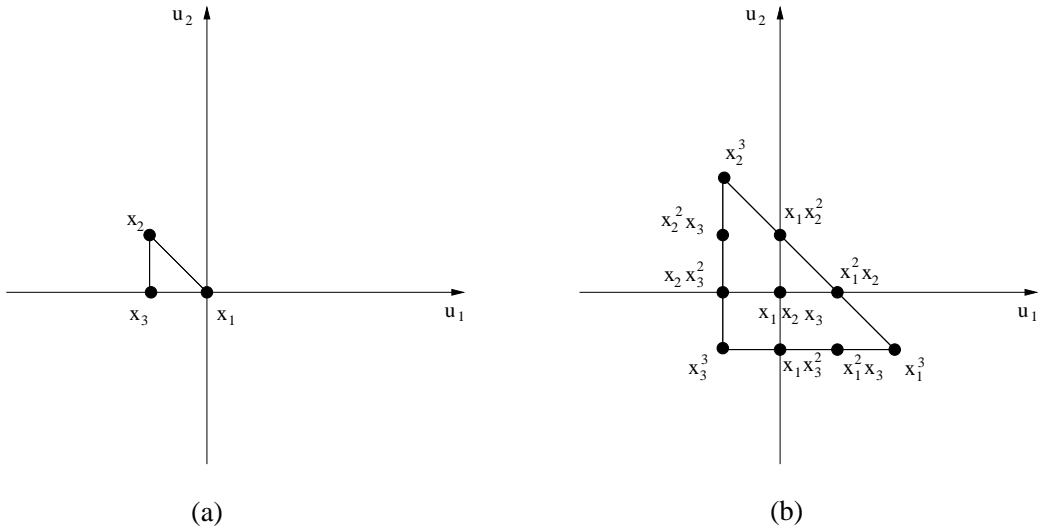} 
\caption{(a) The polytope associated to the line bundle $\mathcal{O}(1)
  \rightarrow \mathbb{P}^2$. (b) The polytope associated to the line bundle $\mathcal{O}(3) \rightarrow \mathbb{P}^2 $.}\label{p2o1o3}
\end{center}
\end{figure}

Consider the line bundle $\mathcal{O}(D_1)$ associated with the divisor $D_1$.
In order to construct its global sections we must first determine the polytope $P_{D_1}$ ($c_1 = 1, c_2 = c_3 = 0$):
\begin{equation}
P_{D_1} = \{ u_1 \geqslant -1 ,\hbox{  } u_2 \geqslant 0 ,\hbox{  } u_1 + u_2 \leqslant 0\}
\end{equation}
Then, using (\ref{sect}), it easy to find the corresponding sections:
\begin{equation}
\{ x_1,\hbox{  } x_2,\hbox{  } x_3 \}
\end{equation}
These are the homogeneous monomials of order one over $\mathbb{P}^2$. Indeed we have just constructed the line bundle 
$\mathcal{O}(1) \rightarrow \mathbb{P}^2$ (see Figure \ref{p2o1o3}). 

Consider as a second example the line bundle $\mathcal{O}(D_1 + D_2 + D_3)$.
In this case the associated polytope is:
\begin{equation}
P_{D_1+D_2+D_3 } = \{ u_1 \geqslant -1 ,\hbox{  } u_2 \geqslant -1 ,\hbox{  } u_1 + u_2 \leqslant 1\}
\end{equation}
Using (\ref{sect}) it is easy to find the corresponding sections:
\begin{equation}
\{ x_1^3,\hbox{  } x_1^2 x_2,\hbox{  } x_1x_2x_3, ...\}
\end{equation}
These are all the homogeneous monomials of degree $3$ over $\mathbb{P}^2$; we have indeed constructed the line bundle 
$\mathcal{O}(3) \rightarrow \mathbb{P}^2$ (see Figure \ref{p2o1o3}).

The examples of polytopes and line bundles presented in this Section are analogous to the ones that we will use in the 
following to characterize the $BPS$ baryonic operators. The only difference (due to the fact that  we are going to consider 
affine toric varieties) is that the polytope $P_D$ will be a non-compact rational convex polyhedron, and the space of 
sections will be infinite dimensional.\\  

\section{BPS D3 brane configurations}\label{susyD3}

In this Section we discuss Beasley's prescription \cite{Beasley:2002xv} 
for determining the BPS 
Hilbert space corresponding to supersymmetric D3 brane configurations. We 
generalize the example of the conifold presented in \cite{Beasley:2002xv}
to the case of a generic toric Calabi-Yau cone.

\subsection{Motivations}\label{motivations}
Consider the supersymmetric background of type $IIB$ supergravity $AdS_5 \times H$ with $H$ a Sasaki-Einstein manifold. This geometry is 
obtained by taking the near horizon geometry of a stack of  $N$ $D3$ on the isolated Goreinstein singularity of a local Calabi-Yau three-fold 
given by the real cone $C(H)$ over the base $H$. The D3 branes fill the four dimensional Minkowski space-time $M_4$ in $M_4 \times C(H)$. 

The dual superconformal field theory is a quiver gauge theory: an $\mathcal{N}=1 $ supersymmetric quantum field theory with gauge group $SU(N_1) \times ...SU(N_k)$ and chiral superfields that transform under the fundamental of a gauge group and the anti-fundamental of another gauge group. 
Due to the presence of $SU(N)$ type groups these theories have generically baryonic like operators inside their spectrum and these are the 
objects we are interested in. 

Let us take the field theory dual to the conifold singularity as a basic example. The theory has gauge group $SU(N) \times SU(N)$ and 
chiral superfields $A_1$, $A_2$ that transform under the fundamental of the first gauge group and under the anti-fundamental of the second one, 
and $B_1$, $B_2$ that transform under the conjugate representation. There exists also a non-abelian global symmetry $SU(2) \times SU(2)$ under which 
the $A$ fields transform as $(2,1)$ and the $B$ as $(1,2)$.  The superpotential
is $W=\epsilon_{ij}\epsilon_{pq} A_i B_p A_j B_q$.  
It is known that this theory has one baryonic charge and that the $A_i$ fields have charge one under this symmetry and the $B_i$ fields have 
charge minus one. Hence one can build the two basic baryonic operators:
\begin{eqnarray}
\label{exbar}
\epsilon^1 _{p_1,...,p_N}  \epsilon_2 ^{k_1,...,k_N} (A_{i_1})^{p_1}_{k_1}... (A_{i_N})^{p_N}_{k_N} = (\det A)_{ ( i_1,...,i_N ) } \nonumber \\
\epsilon^1 _{p_1,...,p_N}  \epsilon_2 ^{k_1,...,k_N} (B_{i_1})^{p_1}_{k_1}... (B_{i_N})^{p_N}_{k_N} = (\det B)_{ ( i_1,...,i_N ) }
\end{eqnarray}
These operators are clearly symmetric in the exchange of the $A_i$ and $B_i$ respectively, and transform under $(N+1,1)$ and $(1,N+1)$ 
representation of $SU(2) \times SU(2)$. The important observation is that these are the baryonic operators with the smallest possible dimension: 
$\Delta _{\det A ,\det B} = N \Delta _{A,B}$.    
One can clearly construct operator charged under the baryonic symmetry with bigger dimension in the following way. Defining the operators 
\cite{Berenstein:2002ke,Beasley:2002xv} \footnote{which are totally symmetric in the $SU(2)\times SU(2)$ indices due to the F-term 
relations $A_i B_p A_j=A_j B_p A_i$, $B_p A_i B_q = B_q A_i B_p$.}
\begin{equation}\label{AAA}
A_{I;J}= A_{i_1}B_{j_1}...A_{i_m}B_{j_m}A_{i_{m+1}}
\end{equation}
the generic type $A$ baryonic operator is:
\begin{equation}
\label{genbarA}
\epsilon^1 _{p_1,...,p_N}  \epsilon_2 ^{k_1,...,k_N} (A_{I_{1};J_{1}})^{p_1}_{k_1}... (A_{I_{N};J_{N}})^{p_N}_{k_N}\, . 
\end{equation}
One can clearly do the same with the type $B$ operators.

Using the tensor relation
\begin{equation}
\label{epsilonss}
\epsilon_{\alpha_1...\alpha_N}\epsilon^{\beta_1...\beta_N}=\delta^{\beta_1}_{[\alpha_1}...\delta^{\beta_N}_{\alpha_N]}\, ,
\end{equation} 
depending on the symmetry of (\ref{genbarA}), one can sometimes factorize the operator in a basic baryon  times operators 
that are neutral under the baryonic charge \cite{Berenstein:2002ke,Beasley:2002xv}. 
It is a notorious fact that the $AdS/CFT$ correspondence maps the basic baryonic operators (\ref{exbar}) to static $D3$ branes 
wrapping specific three cycles of $T^{1,1}$ and minimizing their volumes. The volumes of the $D3$ branes are proportional to 
the dimension of the dual operators in $CFT$. Intuitively, the geometric dual of an operator (\ref{genbarA}) is a fat brane 
wrapping a three cycle, not necessarily of minimal volume, and moving in the $T^{1,1}$ geometry (we will give more rigorous arguments below). 
If we accept this picture the factorizable operators in field theory can be interpreted in the geometric side as the product of gravitons/giant 
gravitons states with a static $D3$ brane wrapped on some cycle, and the non-factorisable ones are interpreted as excitation states of the 
basic $D3$ branes or non-trivial brane configurations. 

What we would like to do is to generalize this picture to a generic conical $CY$ singularity. Using a clever parametrization 
of the possible $D3$ brane $BPS$ configurations in the geometry found in \cite{Mikhailov:2000ya,Beasley:2002xv}, 
we will explain how it is possible to characterize all the baryonic operators in the dual $SCFT$, count them according to their 
charges and extract geometric information regarding the cycles.

\subsection{Supersymmetric D3 brane configurations}\label{supD3}

Consider supersymmetric $D3$ branes wrapping three-cycles 
in $H$. There exists a general characterization of these types of configurations \cite{Mikhailov:2000ya,Beasley:2002xv} that 
relates the $D3$-branes wrapped on $H$ to holomorphic four cycles in $C(H)$. The argument goes as follows. Consider the euclidean 
theory on $\mathbb{R}^4 \times C(H)$. 
It is well known that one $D3$-brane wrapping a holomorphic surface $S$ in $C(H)$ preserves supersymmetry. If we put $N$ $D3$-branes 
on the tip of the cone $C(H)$ and take the near horizon limit the supergravity background becomes $Y_5 \times H$ where $Y_5$ is the 
euclidean version of $AdS_5$. We assume that $S$ intersects $H$ in some three-dimensional cycle $C_3$. 
The $BPS$ $D3$ brane wrapped on $S$ looks like a point in $\mathbb{R}^4$ and like a line in $Y_5$: it becomes a brane wrapped on a 
four-dimensional manifold in $\gamma \times H$ where $\gamma$ is the geodesic in $Y_5$ obtained from the radial direction in $C(H)$. 
Using the $SO(5,1)$ global symmetry of $Y_5$ we can rotate $\gamma$ into any other geodesic in $Y_5$. For this reason when we make the 
Wick rotation to return to Minkowski signature (this procedure preserves supersymmetry) we may assume that $\gamma $ becomes a time-like 
geodesic in $AdS_5$ spacetime. 
In this way we have produced a supersymmetric $D3$ brane wrapped on a three cycle in $H$ which moves along $\gamma$ in $AdS_5$. 
Using the same argument in the opposite direction, we realize also that any supersymmetric $D3$ brane wrapped on $H$ can be 
lifted to a holomorphic surface $S$ in $C(H)$.

 Due to this characterization, we can easily parametrize the classical phase space $\mathcal{M}_{cl}$ of supersymmetric 
$D3$ brane using the space of holomorphic surfaces in $C(H)$ without knowing the explicit metric on the Sasaki-Einstein space $H$ 
(which is generically unknown!).

The previous construction 
characterizes all kind of supersymmetric configurations
of wrapped D3 branes. These include branes wrapping trivial cycles and
stabilized by the combined action of the rotation and the RR flux, which
are called giant gravitons in the literature \cite{McGreevy:2000cw}. Except
for a brief comment on the relation between giant gravitons and dual giant 
gravitons, we will be mostly interested in D3 branes wrapping non trivial
cycles. These correspond to states with non zero baryonic charges in the
dual field theory. The corresponding surface $D$ in $C(H)$ is then a non
trivial divisor, which, modulo subtleties in the definition of the sheaf
${\cal O}(D)$, can be written as the zero locus of a section of ${\cal O}(D)$
\begin{equation}
\chi =0\, \qquad\qquad\qquad \chi\in H^0(X,{\cal O}(D))\label{gen}
\end{equation}

\subsubsection{The toric case}
\begin{figure}
\begin{center}
\includegraphics[scale=0.6]{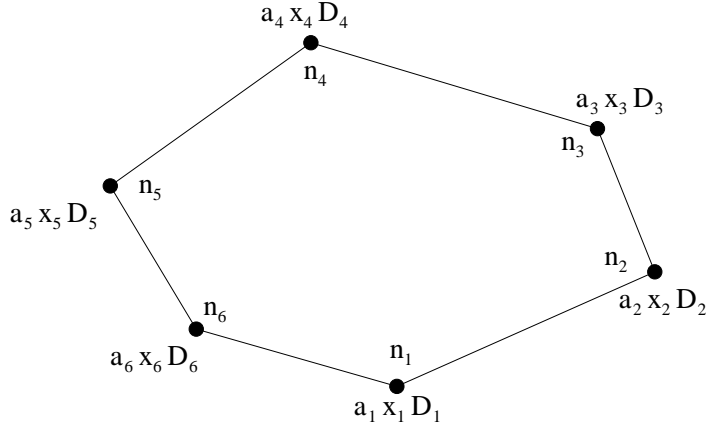} 
\caption{A generic toric diagram with the associated trial charges $a_i$, homogeneous coordinates $x_i$ and divisors $D_i$.}\label{hom}
\end{center}
\end{figure}

The previous discussion was general for arbitrary Calabi-Yau cones $C(H)$.
From now on we will mostly restrict to the case of an affine toric Calabi-Yau cone $C(H)$. For this type of toric manifolds 
the fan $\Sigma$ described in Section \ref{toric} is just a single cone $\sigma$, due to the fact that we are considering a singular affine variety. 
Moreover, the Calabi-Yau nature of the singularity requires that all the generators of the one dimensional cone in $\Sigma(1)$
lie on a plane; this is the case, for example, of the conifold
pictured in Figure \ref{fan}. 
We can then characterize the variety with the convex hull 
of a fixed number of integer points in the plane: the toric diagram (Figure \ref{hom}). 
For toric varieties, the equation for the D3 brane configuration can be written quite explicitly using homogeneous coordinates. 
As explained in Section \ref{toric}, we can associate to every vertex of the toric diagram a global homogeneous coordinate $x_i$. 
Consider a divisor $D=\sum c_i D_i$. All the
supersymmetric configurations of D3 branes corresponding to surfaces linearly
equivalent to $D$ can be written as the zero locus of the generic section
of $H^0 ( V_{\Sigma}, \mathcal{O}_{V_{\Sigma}}(D))$
\begin{equation}\label{holsur}
P(x_1,x_2,...,x_d)\equiv\sum_{m\in P_D \cap M} h_m \chi^m = 0
\end{equation}
As discussed in Section 2, the sections take the form of the
monomials (\ref{sect}) 
$$\chi ^m = \prod _{i=1}^d x_i ^{< m , n_i > + c_i} $$
and there is one such monomial for each
integer point $m\in M$ in the polytope $P_D$ associated with $D$ as 
in equation (\ref{poly})
$$ \{ u \in M_{\mathbb{R}}| <u,n_i >\hbox{  }\geq \hbox{  } - c_i \hbox{  }, \hbox{  }\forall i \in \Sigma(1) \}$$
As already noticed, the $x_i$ are only
defined up to the rescaling (\ref{resc}) but the equation
$P(x_1,...x_d)=0$ makes sense since all monomials have the same charge
under $(\mathbb{C}^*)^{d-3}$ (and under possible discrete orbifold actions). 
Equation (\ref{holsur}) generalizes the familiar
description of hypersurfaces in projective spaces $\mathbb{P}^n$ as zero
locus of homogeneous polynomials. In our case, since we are considering affine varieties, 
the polytope $P_D$ is non-compact and the space of holomorphic global sections is infinite dimensional.
 
We are interested in characterizing the generic supersymmetric $D3$ brane configuration with a fixed baryonic charge. 
We must therefore understand the relation between divisors and baryonic charges: it turns out that there is
a one-to one correspondence between baryonic charges and classes of divisors
modulo the equivalence relation (\ref{eqrel}) \footnote{In a fancy mathematical way, we could say that the 
baryonic charges of a D3 brane configuration are
given by an element of the Chow group $A_2(C(H))$.}. We will understand
this point by analyzing in more detail the $(\mathbb{C}^*)^{d-3}$ action  defined in Section \ref{toric}.  

\subsubsection{The assignment of charges}\label{charges}
To understand the relation between divisors and baryonic charges, 
we must make a digression and recall how one can assign $U(1)$ global charges to the homogeneous coordinates associated 
to a given toric diagram \cite{kru2,tomorrow,aZequiv,proc}.

Non-anomalous $U(1)$ symmetries play a very important role in the dual 
gauge theory and it turns out that we can easily parametrize these global symmetries directly from the toric diagram. 
In a sense, we can associate field theory charges directly to the homogeneous coordinates.

For a background with  horizon $H$, 
we expect $d-1$ global non-anomalous symmetries, 
where $d$ is number of vertices of the toric diagram \footnote{More precisely,
$d$ is the number of integer points along the perimeter of the toric diagram.
Smooth horizons have no integer points along the sides of the toric diagram
except the vertices, and $d$ coincides with the number of vertices. 
Non smooth horizons have sides passing through integer points and these must be
counted in the number $d$.}.
We can count these symmetries by looking at the number of massless vectors in the $AdS$ dual. Since the manifold is toric, 
the metric has three isometries $U(1)^3 \equiv T^3$, which are the real part
of the $(\mathbb{C}^*)^3$ algebraic torus action.
One of these, generated by the Reeb vector, corresponds to the
R-symmetry while the other two give two global flavor symmetries 
in the gauge theory. Other gauge fields in $AdS$ come 
from the reduction of the RR four form on the non-trivial three-cycles
in the horizon manifold $H$, and there are $d-3$ three-cycles in
homology \cite{tomorrow}. On the field theory side, these gauge fields correspond to baryonic 
symmetries. Summarizing, the global non-anomalous symmetries are:
\begin{equation}
U(1)^{d-1}=U(1)^2_F \times U(1)^{d-3}_B
\label{count}
\end{equation}

In this paper we use the fact that these $d-1$ global non-anomalous charges can be parametrized by $d$
parameters $a_1, a_2, \ldots ,a_d$, each associated 
with a vertex of the toric diagram (or a point along an edge), 
satisfying the constraint:
\begin{equation}
\sum_{i=1}^d a_i = 0
\label{sum}
\end{equation}
The $d-3$ baryonic charges are those satisfying the further
constraint \cite{tomorrow}:
\begin{equation}
\sum_{i=1}^d a_i n_i=0
\label{bari}
\end{equation}
where $n_i$ are the vectors of the fan: $n_i=(y_i,z_i,1)$ with
$(y_i,z_i)$ the coordinates of integer points along the perimeter of
the toric diagram. 
The R-symmetries are parametrized with the $a_i$
in a similar way of the other non-baryonic global symmetry, but they satisfy the different constraint
\begin{equation}
\sum_{i=1}^d a_i = 2
\label{sumr}
\end{equation} 
due to the fact that the terms in the superpotential must have
$R$-charges equal to two.

Now that we have assigned trial charges to the vertices of a toric diagram and hence to the homogeneous coordinates $x_i$, 
we can return to the main problem of identifying supersymmetric D3 branes with fixed baryonic charge. 
Comparing equation (\ref{symplq}) with equation (\ref{bari}), we realize that the baryonic charges $a_i$ in 
the dual field theory are the charges $Q^{(a)}_i$ of the action of $(\mathbb{C}^*)^{d-3}$ on the homogeneous coordinates $x_i$. 
We can now assign a baryonic charge to each monomials made with the homogeneous coordinates $x_i$. 
All terms in the equation (\ref{holsur}) corresponding to a D3 brane wrapped on $D$ are global sections $\chi ^m$ of 
the line bundle $\mathcal{O}(\sum_{i=1}^d c_i D_i)$ and they have all the same $d-3$ baryonic charges $B^a=\sum_{i=1}^d Q_i^{(a)} c_i$; 
these are determined only in terms of the $c_i$ defining the corresponding line bundle (see equation (\ref{baryonicaction})). 

Using this fact we can associate a divisor in $C(H)$ to every set of $d-3$ baryonic charges. The procedure is as follows. 
Once we have chosen a specific set of baryonic charges $B^a$ we determine the corresponding $c_i$ using the
relation $B^a=\sum_{i=1}^d c_i Q_i^{(a)}$.  These coefficients define a divisor $D=\sum_{i=1}^d c_i D_i$.  
It is important to observe that, due to equation (\ref{symplq}), the $c_i$ are defined only modulo the equivalence relation 
$c_i \sim c_i + <m,n_i>$ corresponding to the fact that the line bundle $\mathcal{O}(\sum_{i=1}^d c_i D_i)$ is identified only 
modulo the equivalence relations (\ref{eqrel}) $D \sim D + \sum_{i=1}^d <m,n_i> D_i $. We conclude that baryonic charges are 
in one-to-one relation with divisors modulo linear equivalence.


For simplicity, we only considered continuous baryonic charges. In the case
of varieties which are orbifolds, the group $G$ in equation (\ref{Ggroup})
contains a discrete part. 
In the orbifold case, the quantum field theory contains 
baryonic operators with discrete charges. This case can be easily
incorporated in our formalism. 

\subsubsection{The quantization procedure}

Now we want to quantize the classical phase space $\mathcal{M}_{cl}$ using geometric quantization \cite{Wood} following \cite{Beasley:2002xv}.

Considering that $P$ and $\lambda P$ with $\lambda \in \mathbb{C}^*$ vanish on the same locus in $C(H)$, 
it easy to understand that the various distinct surfaces in $C(H)$ correspond to a specific set of $h_m$ 
with $m \in  P_D \cap M $ modulo the equivalence relation $h_m \sim \lambda h_m $: this is a point in the 
infinite dimensional space $\mathbb{CP}^{\infty}$ in which the $h_m$ are the homogeneous coordinates. 
Thus we identify the classical configurations space $\mathcal{M}_{cl}$ of supersymmetric $D3$ brane associated 
to a specific line bundle $\mathcal{O}(D)$ as $\mathbb{CP}^{\infty}$ with homogeneous coordinates $h_m$.
A heuristic way to understand the geometric quantization is the following.
We can think of the $D3$ brane as a particle moving in $\mathcal{M}_{cl}$ and we can associate to it a wave 
function $\Psi$ taking values in some holomorphic line bundle $\mathcal{L}$ over $\mathbb{CP}^{\infty}$.  
The reader should not confuse the line bundle ${\cal L}$, over the classical
phase space $\mathcal{M}_{cl}$ of wrapped D3 branes, with the lines bundles ${\cal O}(D)$, which are defined on $C(H)$.  
Since all the line bundles ${\cal L}$ over a projective space are determined by their degree 
(i.e. they are of the form $\mathcal{O}(\alpha)$) we have only to find the value of $\alpha$. 
This corresponds to the phase picked up by the wavefunction $\Psi$ when the $D3$ brane goes around a closed path in $\mathcal{M}_{cl}$. 
Moving in the phase space $\mathcal{M}_{cl}$ corresponds to moving
 the $D3$ brane in $H$. Remembering that a $D3$ brane couples to the four form field $C_4$ of the supergravity and that 
the backgrounds we are considering are such that $\int_H F_5 =N$, it was argued in \cite{Beasley:2002xv} that the wavefunction 
$\Psi$ picks up the phase $e^{2\pi i N}$ and $\alpha = N$. 
For this reason $\mathcal{L}=\mathcal{O}(N)$ and the global holomorphic sections of this line bundle over 
$\mathbb{CP}^{\infty}$ are the degree $N$ polynomials in the homogeneous coordinates $h_m$. 

Since the $BPS$ wavefunctions are the global holomorphic sections of $\mathcal{L}$, we have that the $BPS$ 
Hilbert space $\mathcal{H}_D$ is spanned by the states: 
\begin{equation}
| h_{m_1}, h_{m_2},...,h_{m_N} > 
\label{bpsstate}
\end{equation} 
This is Beasley's prescription. 

We will make a correspondence 
in the following between $h_m$ and certain operators in the field theory
with one (or more) couple of free gauge indices
\begin{equation}
h_m\, ,\qquad\qquad\qquad (O_m)_{\alpha\beta}\,\, (\rm{or}\,(O_m)_{\alpha_1\beta_1...\alpha_k\beta_k})
\end{equation}
where the $O_m$ are operators with fixed baryonic charge. The generic state
in this sector $| h_{m_1}, h_{m_2},...,h_{m_N} >$ will be identify with a
gauge invariant operator obtained by contracting the $O_m$ with one
(or more) epsilon symbols. The explicit example of the conifold is
discussed in details in \cite{Beasley:2002xv}: the homogeneous coordinates 
with charges $(1,-1,1,-1)$ can be put in one-to-one correspondence with the elementary fields $(A_1,B_1,A_2,B_2)$ 
which have indeed baryonic charge $(1,-1,1,-1)$. The use of the divisor $D_1$ (modulo linear equivalence) allows to
study the BPS states with baryonic charge $+1$. 
It is easy to recognize that the operators $A_{I,J}$ in equation (\ref{AAA}) have baryonic charge $+1$ and are in 
one-to-one correspondence with the sections of ${\cal O}(D_1)$
$$\sum_{m\in P_{D_1}\cap M} h_m \chi^m =h_1 x_1+ h_3 x_3+...$$
The BPS states $|h_{m_1}, h_{m_2},...,h_{m_N}>$
are then realized as all the possible determinants, as  in equation 
(\ref{genbarA}).
 
\subsection{Comments on the relation between giant gravitons and dual giant gravitons}

Among the $\mathcal{O}(D)$ line bundles there is a special one:  the bundle of holomorphic functions $\mathcal{O}$ 
\footnote{Clearly there are other line bundles equivalent to $\mathcal{O}$. For example, since we are considering Calabi-Yau spaces, 
the canonical divisor $K= - \sum_{i=1}^d D_i$ is always trivial and $\mathcal{O} \sim \mathcal{O}(\sum_{i=1}^d D_i)$.}. 
It corresponds to the supersymmetric $D3$ brane configurations wrapped on homologically trivial three cycles $C_3$ in $H$ 
(also called giant gravitons \cite{McGreevy:2000cw}). 

When discussing trivial cycles, we can parameterize holomorphic surfaces just by using the embedding 
coordinates \footnote{These can be also expressed as specific polynomials in the homogeneous coordinates 
such as that their total baryonic charges are zero.}. Our discussion here can be completely general and not 
restricted to the toric case.
Consider the general Calabi-Yau algebraic affine varieties $V$ 
that are cone over some compact base $H$ (they admit at least a $\mathbb{C}^* $ action $V=C(H)$). 
These varieties are the zero locus of a collection of polynomials in some $\mathbb{C}^k$ space. 
We will call the coordinates of the $\mathbb{C}^k$ the embedding coordinates $z_j$, with $j=1,...,k$. 
The coordinate rings $\mathbb{C}[V]$ of the varieties $V$ are the restriction of the polynomials in $\mathbb{C}^k$ 
of arbitrary degree on the variety $V$:   
\begin{equation}
\mathbb{C}[V]= \frac{\mathbb{C}[z_1,...z_k]}{\{p_1,...,p_l\}}= \frac{\mathbb{C}[z_1,...z_k]}{\mathbb{I}[V]}
\label{corring}
\end{equation} 
where $\mathbb{C}[z_1,...,z_k]$ is the $\mathbb{C}$-algebra of polynomials in $k$ variables and $p_j$ are 
the defining equations of the variety $V$.
We are going to consider the completion of the coordinate ring (potentially infinity polynomials) 
whose generic element can be written as the (infinite) polynomial in $\mathbb{C}^k$
\begin{equation}
P(z_1,...z_k) = c + c_i z_i + c_{ij}z_iz_j+...= \sum _I c_I z_I
\label{polCk}
\end{equation} 
restricted by the algebraic relations $\{ p_1=0,...,p_l=0\}$\footnote{ There exist a difference 
between the generic baryonic surface and the mesonic one: the constant $c$. The presence of this 
constant term is necessary to represent the giant gravitons, but if we take for example the constant 
polynomial $P=c$ this of course does not intersect the base $H$ and does not represent a supersymmetric $D3$ brane. 
However it seems that this is not a difficulty for the quantization procedure \cite{Beasley:2002xv,Biswas:2006tj}}. 

At this point Beasley's prescription says that the $BPS$ Hilbert space of the 
giant gravitons $\mathcal{H}_g$ is spanned by the states  
\begin{equation}
| c_{I_1}, c_{I_2},...,c_{I_N} > 
\label{bpsstategiant}
\end{equation} 
These states are holomorphic polynomials of degree $N$ over $\mathcal{M}_{cl}$ and are obviously 
symmetric in the $c_{I_i}$. For this reason we may represent (\ref{bpsstategiant}) as the symmetric product:
\begin{equation}
Sym ( | c_{I_1}> \otimes | c_{I_2}> \otimes ... \otimes |c_{I_N} >) 
\label{bpsstategiantsym}
\end{equation} 
Every element $|c_{I_i} > $ of the symmetric product is a state that 
represents a holomorphic function over the variety $C(H)$. This is easy to understand if one takes the polynomial 
$P(z_1,...z_k)$ and consider the relations among the $c_I$ induced by the radical ring $\mathbb{I}[V]$ (in a sense 
one has to quotient by the relations generated by  $\{ p_1=0,...,p_l=0\}$). For this reason the Hilbert space of giant gravitons is:
\begin{equation}
\mathcal{H}_g =  \bigotimes ^{N \hbox{   } Sym} \mathcal{O}_{C(H)} 
\label{bpsstategiants}
\end{equation} 
Obviously, we could have obtained the same result in the toric case 
by applying the techniques discussed in this Section. Indeed if we put all the $c_i$ equal to zero the 
polytope $P_D$ reduces to the dual cone $\mathcal{C}^*$ of the toric diagram, whose integer points corresponds 
to holomorphic functions on $C(H)$ \cite{MSY2}.

In a recent work \cite{Martelli:2006vh} it was shown that the Hilbert space of a dual giant graviton
\footnote{A dual giant graviton is a D3 brane wrapped on a three-sphere in $AdS_5$} $\mathcal{H}_{dg}$ in 
the background space $AdS_5 \times H$, where $H$ is a generic Sasaki-Einstein manifold, is the space of 
holomorphic functions over the cone $C(H)$. \
At this point it is easy to understand why the counting of $1/2$ $BPS$ states of giant gravitons and 
dual giant gravitons give the same result\cite{Biswas:2006tj,Mandal:2006tk}: the counting of $1/2$ $BPS$ 
mesonic state in field theory. Indeed:
\begin{equation}
\mathcal{H}_g =  \bigotimes ^{N \hbox{   } Sym} \mathcal{H}_{dg} 
\label{bpsstategiantsh}
\end{equation}

\section{Flavor charges of the BPS baryons}\label{bar}
In the previous Section, we discussed supersymmetric $D3$ brane configurations with specific baryonic charge. 
Now we would like to count, in a sector with given baryonic charge, the states with a given 
set of flavor charges $U(1)\times U(1) $ and $R$-charge $U(1)_R$. The generic state of the $BPS$ Hilbert 
space (\ref{bpsstate}) is, by construction, a symmetric product of the single states $|h_m>$. 
These are in a one to one correspondence with the integer points in the polytope $P_D$, which 
correspond to sections $\chi^m$. As familiar in toric geometry \cite{fulton,cox}, a integer point $m\in M$ contains information about
the charges of the $T^3$ torus action, or, in quantum field theory language,
 about the flavor and $R$ charges.

Now it is important to realize that, as already explained in Section \ref{susyD3}, the charges $a_i$ 
that we can assign to the homogeneous coordinates $x_i$ contain information about the baryonic charges 
(we have already taken care of them) but also about the flavor and $R$ charges in the dual field theory. 
If we call $f^k_i$ with $k=1,2$ the two flavor charges and $R_i$ the $R$-charge, 
the section $\chi^m$ has flavor charges (compare equation (\ref{sect})):
\begin{equation}
f^k_{m} = \sum_{i=1}^d (<m,n_i> + c_i)f^k_i
\label{flav}
\end{equation}  
and $R$-charge:
\begin{equation}
R_{m} = \sum_{i=1}^d (<m,n_i> + c_i)R_i
\label{erre}
\end{equation} 

It is possible to refine the last formula. Indeed 
the $R_i$, which  are the R-charges of a set of elementary fields of the gauge theory
\cite{tomorrow,aZequiv} \footnote{The generic elementary field in the gauge
theory has an R-charge which is a linear combination of the $R_i$ \cite{aZequiv}.},
are completely determined by the Reeb vector of $H$ and the vectors $n_i$ defining the toric diagram \cite{MSY}. 
Moreover, it is possible to show that $\sum_{i=1}^d n_i R_i = \frac{2}{3}b $ \cite{aZequiv}, where $b$ specifies how the Reeb vector 
lies inside the  $T^3$ toric fibration. Hence:
\begin{equation}
R_{m} = \frac{2}{3} <m,b> + \sum_{i=1}^dc_iR_i
\label{errebb}
\end{equation} 
This formula generalizes an analogous one for mesonic operators \cite{Butti:2006nk}. 
Indeed if we put all the $c_i$ equal to zero the polytope $P_D$ reduces to the 
dual cone $\mathcal{C}^*$ of the toric diagram \cite{MSY2}. We know  that the elements of the 
mesonic chiral ring of the $CFT$ correspond to integer points in this cone and they have $R$-charge 
equal to $\frac{2}{3}<m,b>$. 
In the case of generic $c_i$, the right most factor of (\ref{errebb}) is in a sense the background $R$ charge: 
the $R$ charge associated to the fields carrying the non-trivial baryonic charges. In the simple example of 
the conifold discussed in subsection \ref{motivations}, formula (\ref{errebb}) applies to the operators (\ref{AAA}) 
where the presence of an extra factor of $A$ takes into account the background charge. In general the R charge (\ref{errebb})
is really what we expect from an operator in field theory that is given by elementary fields with some baryonic charges dressed by ``mesonic insertions''.

The generic baryonic configuration is constructed by specifying $N$ integer points $m_{\rho}$ in the polytope $P_D$. 
 Its $R$ charge $R_B$ is
\begin{equation}
R_{B} = \frac{2}{3} \sum_{\rho = 1}^N <m_{\rho},b> + N \sum_{i=1}^dc_iR_i
\label{errebar}
\end{equation} 
This baryon has $N$ times the baryonic charges of the associated polytope. 
Recalling that at the superconformal fixed point dimension $\Delta$ and $R$-charge of a chiral superfield are related 
by $R = 2 \Delta / 3$, it is easy to realize that the equation (\ref{errebar}) is really what is expected for a baryonic 
object in the dual superconformal field theory. Indeed if we put all the $m_{\rho} $ equal to zero we have 
(this means that we are putting to zero all the mesonic insertions)
\begin{equation}
\Delta _{B} = N \sum_{i=1}^dc_i \Delta _i
\label{basicbar}
\end{equation} 
This formula can be interpreted as follows.
The elementary divisor $D_i$ can be associated with (typically more than one) elementary field in the gauge theory, with R charge $R_i$. 
By taking just one of the $c_i$ different from zero in formula (\ref{basicbar}), we obtain the dimension of a 
baryonic operator in the dual field theory: take a fixed field, compute its determinant and the dimension of 
the operator is $N$ times the dimension of the individual fields. These field operators correspond to $D3$ branes 
wrapped on the basic divisors $D_i$ and are static branes in the $AdS_5 \times H$ background \footnote{The generic configuration of a $D3$ brane wrapped on a three cycle $C_3$ in $H$ is given by a holomorphic section of $\mathcal{O}(D)$ that is a 
non-homogeneous polynomial under the $R$-charge action. For this reason, and holomorphicity, 
it moves around the orbits of the Reeb vector \cite{Mikhailov:2000ya,Beasley:2002xv}. 
Instead the configuration corresponding to the basic baryons is given by the zero locus of a 
homogeneous monomial (therefore, as surface, invariant under the $R$-charge action), and for this reason it is static.}. 
They wrap the three cycles $C_3^i$ obtained by restricting the elementary divisors $D_i$  at $r=1$. 
One can also write the $R_i$ in terms of the volume of the Sasaki-Einstein space $H$ and of the volume of $C_3^i$ \cite{Gubser:1998fp}:
\begin{equation}
R_i = \frac{\pi \hbox{Vol}(C_3^i)}{3 \hbox{Vol}(H)}
\label{errevol}
\end{equation} 
Configurations with
more than one non-zero $c_i$ in equation (\ref{basicbar}) correspond to
basic baryons made with elementary fields whose R-charge is a linear combination of the $R_i$ (see \cite{aZequiv}) 
or just the product of basic baryons. 

The generic baryonic configuration has $N$ times the $R$-charge and the global charges of the basic baryons 
(static branes which minimize the volume in a given homology class) plus the charges given by the 
fattening  and the motion of the three cycle inside the geometry (the mesonic fluctuations on the $D3$ 
brane or ``mesonic insertions'' in the basic baryonic operators in field theory). It is important to notice 
that the BPS operators do not necessarily factorize in a product of basic baryons times mesons 
\footnote{In the simple case of the conifold this is due to the presence of two fields $A_i$ with the same gauge 
indices; only baryons symmetrized in the indices $i$ factorize \cite{Berenstein:2002ke,Beasley:2002xv}. 
In more general toric quiver gauge theories it is possible to find different strings of elementary fields with the same
baryonic charge connecting a given pair of gauge groups; their existence 
prevents the generic baryons from being factorizable.}.

\subsection{Setting the counting problem}
In Section \ref{counting} we will count the baryonic states of the theory with given baryonic charges (polytope $P_D$) 
according to their $R$ and flavor charges. 
Right now we understand the space of classical supersymmetric $D3$ brane configurations $\mathcal{N}$ as a 
direct sum of holomorphic line bundles over the variety $C(H)$:
\begin{equation}
\mathcal{N} = \bigoplus_{c_i \sim c_i + <m,n_i> } \mathcal{O}\Big(\sum_i^d c_i D_i \Big)
\label{dirs}
\end{equation} 
where the $c_i$ specify the baryonic charges. We have just decomposed the
space $\mathcal{N}$ into sectors according to the grading given by the baryonic
symmetry. Geometrically, this is just the decomposition of the homogeneous
coordinate ring of the toric variety under the grading given by the action
of $(\mathbb{C}^*)^{d-3}$. Now, we want to introduce a further grading.
Inside every line bundle there are configurations with different flavor and $R$ charges. 

Once specified the baryonic charges, the Hilbert space of BPS
operators is the $N$ order symmetric product of the corresponding line bundle. 
Hence the $1/2$ $BPS$ Hilbert space is also decomposed as:
\begin{equation}
\mathcal{H} = \bigoplus_{c_i \sim c_i + <m,n_i> } \mathcal{H}_{D}
\label{dirsH}
\end{equation} 
We would like to count the baryonic operators of the dual $SCFT$ with a given set of flavor and $R$ charges. 
We can divide this procedure into three steps:
\begin{itemize}
\item{find a way to count the global sections of a given holomorphic line bundle (a baryonic partition function $Z_D$);}
\item{write the total partition function for the $N$-times symmetric product of the polytope $P_D$ (the partition function $Z_{D,N}$). 
This corresponds to find how  many combinations there are with the same global charges $a_B^k$ 
(with $k=1,2$ for the flavor charges and $k=3$ for the $R$ charge ) for a given baryonic state: the possible set of $m_{\rho}$ such that: 
\begin{equation}
a_B^k - N \sum_{i=1}^d c_i a_i^k = \sum_{m_{\rho} \in P_D \cap M}\sum_{i=1}^d <m_{\rho},n_i> a_i^k\, .
\label{flavBN}
\end{equation}
}  
\item{write the complete $BPS$ partition function of the field theory by summing over
all sectors with different baryonic charges. 
Eventually
we would also like 
to write the complete $BPS$ partition function of the field theory including all the $d$ charges at a time: 
$d-3$ baryonic, $2$ flavor and $1$ $R$ charges \cite{ADAZ}.}
\end{itemize} 
In the following Sections, we will solve completely the first two steps. 
The third step is complicated by various facts. First of all 
the correspondence between the
homogeneous coordinates and fields carrying the same $U(1)$ charges is not one
to one. From the gravity side of the $AdS/CFT$ correspondence one can explain
this fact as follow \cite{tomorrow}. The open strings attached to a $D3$
brane wrapped on the non-trivial three cycles corresponding to the basic
baryons in the dual field theory have in general many supersymmetric vacuum
states. This multiplicity of vacua corresponds to the fact that generically
the first homotopy group of the three cycles $\pi_1(C_3)$ is non-trivial and
one can turn on a locally flat connection with non-trivial Wilson lines. The
different Wilson lines give the different open string vacua and these are
associated with different elementary fields $X_{ij}$ (giving rise to basic
baryons $\det X_{ij}$ with the same global charges). One has then to include
non trivial multiplicities for the $Z_{D,N}$ when computing the complete BPS 
partition function. Moreover one should pay particular attention to the
sectors with higher baryonic charge. All these issues and the
determination of the partition function depending on all
the $d$ charges will be discussed in forthcoming publications \cite{ADAZ}.

\section{Comparison with the field theory side}  
At this point it is probably worthwhile to make a more straight contact with 
the field theory. This is possible at least for all toric CY because the
 dual quiver
gauge theory is known \cite{dimers,rhombi,mirror}. In this paper we will
mainly focus on the partition function $Z_D$ for supersymmetric D3 brane 
configurations. In forthcoming papers \cite{ADAZ} we will show how to
compute the partition function for the chiral ring and how to compare with
the full set of BPS gauge invariant operators. In this Section we show that,
for a selected class of polytopes $P_D$, there is a simple correspondence
between sections of the line bundle $D$ and operators in the gauge theory.
  
\begin{figure}
\begin{center}
\includegraphics[scale=0.6]{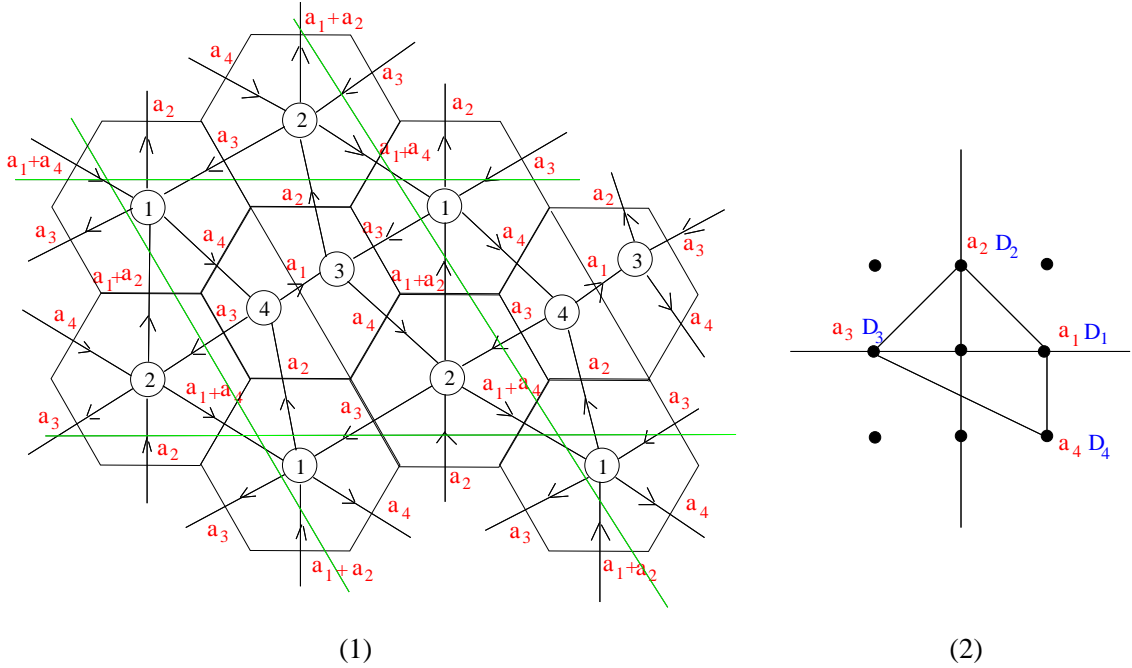} 
\caption{(1) Dimer configuration for the field theory dual to $C(Y^{2,1})$ with a given assignment of 
charges $a_i$ and the orientation given by the arrows connecting the gauge groups. We have drawn in green the bounds of the basic cell. 
For notational simplicity we have not indicated with different colors the
vertices; the dimer is a bipartite graph and this determines an orientation.
(2) Toric diagram for the singularity $C(Y^{2,1})$.}\label{barydp}
\end{center}
\end{figure}

The gauge theory dual to a given toric singularity is completely identified by the \emph{dimer configuration}, 
or \emph{brane tiling} (Figure \ref{barydp}) \cite{dimers,rhombi,mirror}.
This is a bipartite graph drawn on a torus $T^2$: it has an equal number of white and 
black vertices and links connect only vertices of different colors.
In the dimer the faces represent $SU(N)$ gauge groups, oriented
links represent chiral bifundamental multiplets and nodes represent
the superpotential: the trace of the
product of chiral fields around a node gives a superpotential 
term with sign + or - according to whether the vertex is a white one or a black one. 
By applying Seiberg dualities to a quiver gauge theory we can obtain
different quivers that flow in the IR to the same CFT: to a toric
diagram we can associate different quivers/dimers describing the same
physics. It turns out that one can always find phases where all the 
gauge groups have the same number of colors; these are called 
\emph{toric phases}. Seiberg dualities keep constant the number of
gauge groups $F$, but may change the number of fields $E$, and
therefore the number of superpotential terms $V=E-F$. The toric phases having the minimal set of fields are called \emph{minimal toric phases}.

There is a general recipe for assigning baryonic, flavors and R charges to
the elementary fields for a minimal toric phase of the $CFT$ \cite{kru2,tomorrow,aZequiv,proc}.
As described in Section (\ref{charges}), we can parameterize all charges
with $d$ numbers $a_i$, $i=1,...,d$ associated with the 
vertices of the toric diagram subject to the constraint (\ref{sum})
in case of global symmetries and (\ref{sumr}) in case of R symmetries.
Every elementary field can be associated with a brane wrapping a
particular divisor 
\begin{equation}
D_{i+1}+D_{i+2}+ \ldots D_{j}
\end{equation}
and has charge $a_{i+1}+a_{i+2}+ \ldots a_{j}$ \footnote{Call $C$ the set of all the unordered pairs of vectors in the $(p,q)$ web (the $(p,q)$ web 
is the set of vectors $v_i$ perpendicular to the edges of the toric diagram and with the same length as the corresponding edge); 
label an element of $C$ with the ordered indexes $(i,j)$, with the convention that  
the vector $v_i$ can be rotated to $v_j$ in the counter-clockwise direction
with an angle $\leq 180^o$. With our conventions $|\langle v_i, v_j \rangle|=\langle v_i, v_j \rangle$, 
where with $\langle \hbox{ },\hbox{ } \rangle$ we mean the determinant of the $2 \times 2$ matrix. One can associate with any element of $C$ the divisor 
$D_{i+1}+D_{i+2}+ \ldots D_{j}$ 
and a type of chiral field in the field theory with multiplicity $\langle v_i, v_j \rangle$ 
and global charge equal to $a_{i+1}+a_{i+2}+ \ldots a_{j}$ \cite{aZequiv}. 
The indexes $i$, $j$ are always understood to be defined modulo $d$.}. Various fields have the same
charge; as mentioned in the last part of the Section \ref{bar}, this multiplicity is due to the non-trivial homotopy of the corresponding cycles.
Explicit methods for computing the charge of each link in the dimer are
given 
in \cite{tomorrow,aZequiv}. The specific example of $H=Y^{2,1}$ is reported
in Figure \ref{barydp}. 

With this machinery in our hands we can analyze the the field theory operators 
corresponding to the $D3$ brane states analyzed in the previous Sections. 
The first thing to understand is the map between the section $\chi^m$ of the 
line bundle we are considering and the field theory operators. The case of 
the trivial line bundle is well known: the corresponding polytope
is the cone of holomorphic functions which are in one-to-one correspondence
with mesonic operators. The latter are just closed loops in the quiver.
It is possible to construct a map between closed loops in the quiver 
and points in the cone of holomorphic functions; it can be
shown that closed loops mapped to the same point in the cone correspond
to mesons that are $F$-term equivalent \cite{hanherve,ago}.
  
We would like to do the same with open paths in the dimer. In particular
we would like to associate to every point in the polytope $P_D$ a sequence of 
contractions of elementary fields modulo $F$-term equivalence. This is indeed
possible for a particular class of polytopes which we now describe. 

Let us start by studying open paths in the dimer.
Take two gauge group $U$, $V$ in the dimer and draw an oriented path $P$ connecting them 
(an oriented path in the dimer is a sequence of chiral fields oriented in the same way and with the gauge indices contracted along the path). 
The global charges of $P$ is the sum of the charges of the fields contained in $P$ and can be schematically written as:
\begin{equation}\label{P}
\sum_{i=1}^d c_i a_i
\end{equation}
with some integers $c_i$. Draw another oriented path $Q$ connecting the same gauge groups. 
Consider now the closed non-oriented path $Q-P$; as explained in \cite{ago} the charges 
for a generic non-oriented closed path can be written as \footnote{For non-oriented paths one has 
to sum the charges of the fields with the same orientation and subtract the charges 
of the fields with the opposite orientation.}:
\begin{equation}\label{P-Qa}
\sum_{i=1}^d <m,n_i> a_i 
\end{equation}
with $m$ a three dimensional integer vector.
Hence the charges for a generic path $Q$ connecting two gauge groups are:
\begin{equation}\label{P-Qb}
\sum_{i=1}^d (<m,n_i> + c_i) a_i\, .
\end{equation}
 Because the path $Q$ is oriented we have just added the charges along it and the coefficients of the $a_i$ are all positive:
\begin{equation}\label{P-Qc}
<m,n_i> + c_i \geqslant 0\, .
\end{equation}
We have the freedom to change $P$; this means that the $c_i$ are only 
defined up to the equivalence relation $c_i \sim c_i + <m,n_i>$.
Observe that (\ref{P-Qc}) is the same condition on the exponents for the 
homogeneous coordinates $x_i$ in the global sections $\chi^m$ (\ref{poly}), 
and the equivalence relation $c_i \sim c_i + <m,n_i>$ corresponds to the equivalence relation 
on the divisors $D \sim D + \sum_{i=1}^d <m,n_i> D_i$. 
Hence we realize that to every path connecting a pair of gauge groups we can assign 
a point in a polytope associated with the divisor  $\sum_i c_i D_i$  modulo 
linear equivalence. In particular, all operators associated with open paths
between two gauge groups $(U,V)$ have the same baryonic charge, as it can
be independently checked.
 
Now that we have a concrete map between the paths in the dimer and the integer 
points in the polytope we have to show that this map is well defined. Namely 
we have to show that we map $F$-term equivalent operators to the same point in the polytope and 
that to a point in the polytope corresponds only one operator in field theory modulo $F$ terms relations (the injectivity of the map). 
The first step is easy to demonstrate: paths that are $F$-term equivalents have the 
same set of $U(1)$ charges and are mapped to the same point $m$ in the polytope $P_D$. 
Conversely if paths connecting two gauge groups are mapped to the same point $m$ it means 
that they have the same global charges. The path $P-P'$ is then
a closed unoriented path with charge 0. As shown in \cite{hanherve,ago}
$P$ and $P'$ are then homotopically equivalent \footnote{ Indeed it is possible to 
show that $m_1$ and $m_2$ of a closed path are its homotopy numbers around the dimer. }.  
Now we can use the Lemma $5.3.1$ in \cite{hanherve} that says: ``in a consistent tiling, 
paths with the same $R$-charge are $F$-term equivalent if and only if they are homotopic'' 
to conclude that paths mapped to the same point $m$ in $P_D$ are $F$-term equivalent. Surjectivity of the map is more difficult to prove, exactly 
as in the case
of closed loops \cite{hanherve,ago}, but it is expected
to hold in all relevant cases.


In particular we will apply the previous discussion to the case of neighbouring
gauge groups $(U,V)$ connected by one elementary field. If the charge of
the field is $a_{i+1}+...+a_j$ we are dealing with the sections of the line
bundle ${\cal O}(D_{i+1}+...+D_j)$
$$\sum_{m\in P_{D_{i+1}+...+D_j}} h_m \chi^m =h\,  x_{i+1}...x_j+...$$
The section $x_{i+1}...x_j$ will correspond to the elementary field itself
while all other sections $\chi^{m_j}$ will correspond to operators with
two free gauge indices $(O_m)_{\alpha\beta}$ under $U$ and $V$ which correspond to open paths from $U$ to $V$.  
The proposal for finding the gauge invariant operator dual to the $BPS$ state $|h_{m_1},...,h_{m_N}>$ 
is then the following\footnote{ This is just a simple generalization of the one in \cite{Beasley:2002xv}.}. 
We associate to every $h_{m_j}$, with section $\chi^{m_j}$, 
an operator in field theory with two free gauge indices 
in the way we have just described (from now on we will call these paths the building blocks of the baryons).  
Then we construct a gauge invariant operator by contracting all the $N$ free indices of one gauge group with its epsilon tensor and all the $N$ free indices of the other gauge group with its own epsilon. 
The field theory operator we have just constructed has clearly the same global charges 
of the corresponding state in the string theory side and due to the epsilon contractions 
is symmetric in the permutation of the field theory building blocks like the string theory state. This generalizes Equation (\ref{genbarA}) to the case of
a generic field in a toric quiver.  
By abuse of language, we can say that
we have considered all the single determinants that we can make with
indices in $(U,V)$.

As already mentioned, D3 branes wrapped
on three cycles in $H$ come with a multiplicity which is given by the
non trivial homotopy of the three cycle. On the field theory side, this
corresponds to the fact there is a multiplicity of elementary fields with
the same charge. Therefore a polytope $P_D$
is generically associated to various 
different pairs of gauge groups $(U^a,V^a)$, $a=1,...\#_D$. 
For this reason we say that the polytope $P_D$ has a multiplicity $\#_D$. 
This implies that there is an isomorphism between the set of 
open paths (modulo $F$-terms) connecting the different pairs $(U^a,V^a)$. 
Similarly, the single determinant baryonic operators
constructed as above from different pairs $(U^a,V^a)$ come isomorphically
from the point of view of the counting problem.




Obviously, the baryonic operators we have constructed are just a subset
of the chiral ring of the toric quiver gauge theory. They correspond to
possible single determinants that we can construct. In the case of greater
baryonic charge (products of determinants) the relation between points
in the polytope $P_D$ and operators is less manifest and it
will be discussed in section \ref{comments}. 

As an example of this construction, we now discuss the baryonic building blocks associated with a line bundle over the Calabi-Yau cone $C(Y^{2,1})$.
\subsection{Building blocks for $\mathcal{O}(D_3) \rightarrow C(Y^{2,1})$}
Let us explain the map between the homogeneous coordinates and field theory operators in a simple example: $C(Y^{2,1})$. 
The cone over $Y^{2,1}$ has four divisors with three equivalence relations (see Figure \ref{barydp}) 
and hence we have the assignment:
\begin{equation}\label{divy21}
D_1= 3D \qquad  D_2 = -2D \qquad  D_3=D \qquad D_4= -2D
\end{equation}
We want to construct the building blocks of the $BPS$ operators with baryonic charge equal to one. 
Because the $(\mathbb{C}^*)^{d-3}=\mathbb{C}^* $ action is specified by the charge:
\begin{equation}\label{Qy21}
Q=(+3,-2,+1,-2)
\end{equation}
we choose $c_3=1$ and $c_1=c_2=c_4=0$. Hence:
\begin{equation}\label{PD3}
P_{D_3}=\{ m \epsilon M | m_1+m_3 \geq 0, m_2 + m_3 \geq 0, -m_1 + m_3 \geq -1, m_1 - m_2 + m_3 \geq 0 \}
\end{equation}
We can now easily construct the sections of the corresponding line bundle $\mathcal{O}(D_3)$ 
and try to match these with the $BPS$ operators, which are just the open paths in the dimer 
(see Figure \ref{barydp}) with the same trial charges $a_i$ of the polynomial in the homogeneous variables.
Looking at the dimer of $Y^{2,1}$ we immediately realize that there are three distinct 
pairs of gauge groups with charge $\sum_{i=1}^dc_i a_i = a_3$: $(2,1)$, $(4,2)$, $(1,3)$. 
Hence the multiplicity of $P_{D_3}$ is $\#_{D_3} = 3$ and for every point in the polytope 
we have three different operators in the field theory side, corresponding to paths in the 
dimer connecting the three different pairs of gauge groups. In Table $1$ we match the 
sections in the geometry side with the operators in the field theory side for few points 
in the polytope $P_{D_3}$.
\\
\\

\begin{small}
\begin{tabular}{|c|c|c|lll|}
\hline
$(m_1,m_2,m_3)$ & sections & charges  & \verb| | $(2,1)$ & \verb| | $(4,2)$ & \verb| | $(1,3)$ \\ \hline
(0,0,0) & $x_3$ & $a_3$ & $X_{21}^{dl}=Y$ & $X_{42}^{dl}=Y$ & $X_{13}^{dl}=Y$ \\ \hline
(1,0,0) & $x_1 x_4$  & $a_1 + a_4$  & $X_{21}^{dr}=V$ & $X_{43}^{ur}X_{32}^{dr}=ZU$ &  $X_{14}^{dr}X_{43}^{ur}=UZ$ \\ \hline
(1,1,0) & $x_1 x_2$  & $a_1 + a_2$  & $X_{21}^{u}=V$ & $X_{43}^{ur}X_{32}^{u}=ZU$ &  $X_{14}^{u}X_{43}^{ur}=UZ$  \\ \hline
(-1,0,1) & $x_2 x_3^3$  & $a_2 + 3 a_3$  & $X_{21}^{dl}X_{14}^{u}X_{42}^{dl}X_{21}^{dl}$ & $X_{42}^{dl}X_{21}^{dl}X_{14}^{u}X_{42}^{dl}$ &  $X_{13}^{dl}X_{32}^{u}X_{21}^{dl}X_{13}^{dl}$  \\ 
\, & \, & \,  & $=YUYY$ & $=YYUY$ &  $=YUYY$  \\ \hline
(-1,-1,1) & $x_3^3 x_4$  & $3 a_3 + a_4 $  & $X_{21}^{dl}X_{13}^{dl}X_{32}^{dr}X_{21}^{dl}$ & $X_{42}^{dl}X_{21}^{dl}X_{14}^{dr}X_{42}^{dl}$ &  $X_{13}^{dl}X_{32}^{dr}X_{21}^{dl}X_{13}^{dl}$  \\ 
\, & \,  & \,  & $=YYUY$ & $=YYUY$ &  $=YUYY$  \\ \hline
...& ...& ...& ... & ... & ... \\ \hline
\end{tabular}
\vskip0.5truecm
Table 1: Few sections of $\mathcal{O}(D_3)$ and the corresponding field theory operators 
of baryon number $1$. We write: the point $m$ in the polytope; the corresponding section 
($x_i$ are the homogeneous coordinates ); its charges; the three corresponding gauge operators 
($X$ are the fundamental fields): we used the label $u$, $l$, $d$, $r$ for $up$, $left$, $down$, $right$, 
to specify the field direction and, for comparison with the literature, the field is also written 
using the notations commonly adopted for $Y^{p,q}$ \cite{benvenuti}.
\end{small}
\vspace{0.5cm} 

One can observe, by looking at the dimer (Figure \ref{barydp}), 
that in the fourth and fifth lines of Table $1$ one can assign different operators to the same section, 
but it is easy to check that these are related by $F$-term equations. Hence in this simple case 
the correspondence between geometry and field theory is manifest. 

The gauge invariant operators with baryonic charge $N$ 
are obtained by taking all the operators connecting 
the same gauge groups and by contracting the $N$ free indices of one 
gauge group with its epsilon and the $N$ free indices of the other gauge group 
with its own epsilon. 
All operators come in triples with the same quantum numbers.

\subsection{Comments on the general correspondence}\label{comments}

In the case of a generic polytope $P_D$, not associated with elementary
fields, the correspondence between sections of the line bundle ${\cal O}(D)$
and operators is less manifest. The reason is that
we are dealing with higher baryonic charges and the corresponding gauge
invariant operators are generically products of determinants.

\begin{figure}
\begin{center}
\includegraphics[scale=0.5]{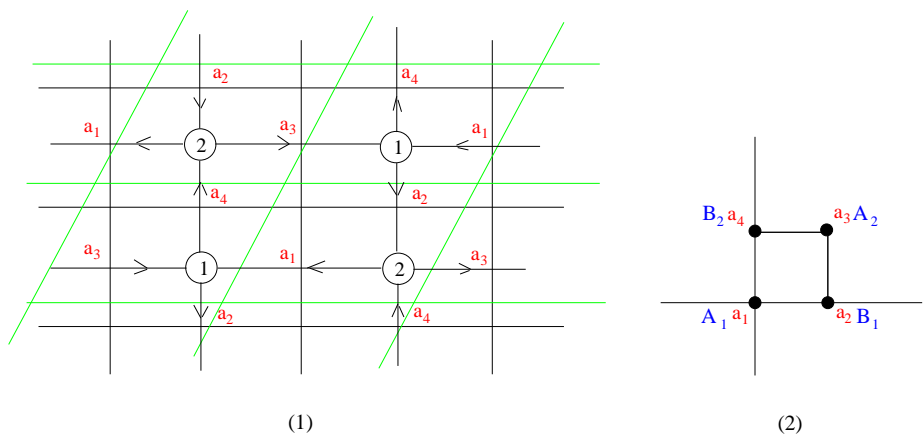} 
\caption{(1) Dimer configuration for the field theory dual to $C(T^{1,1})$
  with a given assignment of charges $a_i$ and the orientation given by the
  arrows linking the gauge groups. We have drawn in green the bounds of the
  basic cell. (2) Toric diagram for the singularity $C(T^{1,1})$.}\label{baryt}
\end{center}
\end{figure}

Let us consider as an example the case of the conifold. Suppose we want to 
study the polytope $P_{2 D_1}$ which corresponds to 
classify the $BPS$ operators with baryonic charges equal to $2N$.
In field theory we certainly have baryonic operators with charge $2N$, 
for example $\det A_1\cdot \det A_1 $. 
Clearly all the products of two baryonic operators 
with baryon number $N$ give a baryonic operator with baryon number $2N$. 
In a sense in the conifold all the operators in sectors with baryonic charge 
with absolute value bigger than one are factorized \cite{Berenstein:2002ke,Beasley:2002xv}. However we cannot find a simple prescription 
for relating sections  of $P_{2D_1}$ to paths in the dimer. Certainly
we can not find a single path in the dimer (Figure \ref{baryt}) 
connecting the two gauge groups with  charge $2 a_1$. 

One could speculate that the prescription valid for basic polytopes
has to be generalized by allowing the use of paths and multipaths.
For example, to the section $\chi^m$ in the polytope $P_{2D_1}$ for the conifold 
we could assign two paths connecting the two gauge groups with charges $\sum_{i=1}^d <m^{(1)},n_i> + a_1$ 
and  $\sum_{i=1}^d <m^{(2)},n_i> + a_1$ with $m=m^{(1)}+m^{(2)}$ and therefore a building block
consisting of two operators $(O_{m^{(1)}})_{\alpha_1\beta_1} (O_{m^{(2)}})_{\alpha_2\beta_2}$. We should now construct the related gauge invariant operators. 
Out of these building blocks we cannot construct a single determinant because 
we don't have an epsilon symbol with $2N$ indices, but we can easily construct 
a product operator using four epsilons. 
We expect, based on Beasley's prescription, a one to one correspondence 
between the points in the $N$ times symmetric product of the polytope $P_{2 D_1}$ 
and the baryonic operators with baryonic charge $+2$ in field theory. 
Naively, it would seem that, with the procedure described above, we have found many more operators. 
Indeed the procedure was plagued by two ambiguities: in the construction of the 
building blocks, it is possible to find more than a pair of paths corresponding
to the same $m$ (and thus the same $U(1)$ charges) that are not $F$-term equivalent; 
in the construction of the gauge invariants we have the ambiguity on how to distribute 
the operators between the two determinants. 
The interesting fact is that these ambiguities seem to disappear when we
consider the final results for gauge invariant operators, due to the $F$-term relations and the properties of the epsilon symbol. 
One can indeed verify, at least in the case of $P_{2 D_1}$ and for 
various values of $N$, 
there is exactly a one to one correspondence between the points in the $N$ times symmetric 
product of the polytope and the baryonic operators in field theory. 
It would be interesting to understand if this kind of prescription can be 
made rigorous.

The ambiguity in making a correspondence between sections of the polytope
and operators is expected and it is not particularly problematic.
The correct correspondence is between the states $|h_{m_1},...,h_{m_N}>$ and 
baryonic gauge invariant operators. 
The sections in the geometry are not states 
of the string theory and the paths/operators are not gauge invariant operators. What the $AdS/CFT$ 
correspondence tells us is that there exists a 
one to one relation between states in string theory 
and gauge invariant operators in field theory, and this is a one to one 
relation between 
the points in the $N$-fold symmetric product of a given polytope and 
the full set of gauge invariant operators with given baryonic charge.

The comparison with field theory should be then done as follows. One
computes the partition functions $Z_{D,N}$ of the $N$-fold symmetrized
product of the polytope $P_D$ and compares it with all the gauge invariant 
operators in the sector of the Hilbert space with given baryonic charge.
We have explicitly done it for the conifold for the first few values of $N$
and the operators with lower dimension. In forthcoming papers \cite{ADAZ} 
we will actually resum the partition functions $Z_{D,N}$ and we will write
the complete partition function for the chiral ring of the conifold and other
selected examples; we will compare the result with the dual field theory
finding perfect agreement.

The issues of multiplicities that we already found in the case of polytopes
associated with elementary fields persists for generic polytopes. Its
complete understanding is of utmost importance for writing a full partition
function for the chiral ring \cite{ADAZ}.

\section{Counting BPS baryonic operators}\label{counting}
 
In this Section, as promised, we count the number of BPS baryonic operators in
the sector of the Hilbert space $\cal{H}_D$, associated with a divisor $D$.
All operators in $\cal{H}_D$ have fixed baryonic charges. Their number
is obviously infinite, but, as we will show, the number of operators with
given charge $m\in T^3$ under the torus action is finite. It thus 
makes sense to write a partition function $Z_{D,N}$ for the BPS baryonic operators 
weighted by a $T^3$ charge $q=(q_1,q_2,q_3)$. $Z_{D,N}$ will be a polynomial in the $q_i$ 
such that to every monomial $n \hbox{ } q_1^{m_1} q_2^{m_2} q_3^{m_3}$ we associate $n$ $BPS$ 
D3 brane states with the R-charge and the two flavor charges
parametrized by $\sum_{i=1}^d (<m,n_i> a_i + N c_i a_i)$. 

The computation of the weighted partition function is done in two steps.
We first compute a weighted partition function $Z_D$, 
or character, counting the sections of ${\cal O}(D)$; these correspond to the $h_m$ which are the elementary constituents of the baryons. 
In a second time, we determine the total partition function $Z_{D,N}$ for
the states $|h_{m_1}...h_{m_N}>$ in $\cal{H}_D$.

\subsection{The character $Z_D$}\label{riemannroch}
We want to resum the character, or weighted partition function, 
\begin{equation}
Z_D = \rm{Tr} \{ q | H^0(X,{\cal O}(D)\} = \sum_{m\in P_D  \cap  M} q^m
\label{character}
\end{equation}   
counting the integer points in the polytope $P_D$ weighted with their
charge under the $T^3$ torus action. 

In the trivial case ${\cal O}(D)\sim {\cal O}$, $Z_D$ is
just the partition function for holomorphic functions discussed in
\cite{MSY2,Benvenuti:2006qr}, which can be computed using the Atyah-Singer index
theorem \cite{MSY2}. Here we show how to extend this method to the
computation of $Z_D$ for a generic divisor $D$. 

Suppose that we
have a smooth variety and a line bundle ${\cal O}(D)$ with a holomorphic action of $T^k$ (with $k=1,2,3$ and $k=3$ is the toric case). 
Suppose also that the higher dimensional cohomology of the line bundle vanishes, $H^{i}(X,{\cal O}(D))=0$, for $i\ge1$. 
The character (\ref{character}) then coincides with the Leftschetz number
\begin{equation}
\chi(q,D) = \sum_{p=0}^3 (-1)^p  \rm{Tr} \{ q | H^{p}(X,{\cal O}(D))\}
\end{equation}
which can be computed using the index theorem \cite{AS}: 
we can indeed write $\chi(q,D)$
as a sum of integrals of characteristic classes over
the fixed locus of the $T^k$ action. In this paper, we will only consider 
cases where $T^k$ has isolated fixed points $P_I$. The general case can be 
handled in a similar way.
In the case of isolated fixed points,
the general cohomological formula\footnote{Equivariant Riemann-Roch, or the Lefschetz fixed point formula, reads
\begin{equation}
\chi(q,D) = \sum_{F_i} \int_{F_i} \frac {{\rm Todd}(F_i) Ch^q (D)}{\prod_{\lambda}(1-q^{m_\lambda^i} e^{-x_\lambda})}
\end{equation}
where $F_i$ are the set of points, lines and surfaces which are fixed by
the action of $q\in T^k$, ${\rm Todd}(F)$ is the Todd class ${\rm Todd}(F) =1 + c_1(F)+...$ and, on a fixed locus, $Ch^q(D)= q^{m^0}e^{c_1(D)}$ where $m^0$ is the weight of the $T^k$ action. The normal bundle $N_i$ of each fixed submanifold $F_i$ has been splitted in
line bundles; $x_\lambda$ are the basic characters and $m_\lambda^i$ the weights of the $q$ action on the line bundles.} considerably simplifies and can be 
computed by linearizing the $T^k$ action near the fixed points. The
linearized action can be analysed as follows.  
Since $P_I$ is a fixed point, the
group $T^k$ acts linearly on the normal (=tangent) space at $P_I$,
$TX_{P_I}\sim \mathbb{C}^3$. 
The tangent space will split into three one dimensional
representations $TX_{P_I}=\sum_{\lambda=1}^3 L^\lambda$ of the abelian group 
$T^k$. We denote the
corresponding weights for the $q$ action with $m_I^\lambda, \lambda=1,2,3$.
Denote also with $m_I^0$ the weight of the action of $q$ on the $\mathbb{C}$ 
fiber of the line bundle ${\cal O}(D)$ over $P_I$. The equivariant Riemann-Roch   formula expresses the Leftschetz number as a sum over the
fixed points
\begin{equation}
\chi(q,D) = \sum_{P_I} \frac{q^{m_I^0}}{\prod_{\lambda=1}^3 (1-q^{m_I^\lambda})}
\end{equation}

We would like to apply the index theorem to our Calabi-Yau cone.
Unfortunately, $X=C(H)$ is not smooth and a generic element of $T^k$ has a fixed point at the apex of the cone, which is exactly the singular point.
To use Riemann-Roch we need to resolve the cone $X$ to a smooth variety 
$\tilde X$
and to find a line bundle ${\cal O}(\tilde D)$ on it with the following two
properties: i) it has the same space of sections, $H^0 (\tilde X,{\cal O}(\tilde D)) = H^0(X,{\cal O}(D))$, ii) it has vanishing higher cohomology
$H^{i}(\tilde X,{\cal O}(\tilde D))=0, i\ge 1$.

Notice that the previous discussion was general and apply to all Sasaki-Einstein manifolds $H$. It gives 
a possible prescription for computing $Z_D$ even in the non toric case.
In the following we will consider the case of toric cones where the
resolution $\tilde X$ and the divisor $\tilde D$ can be explicitly found.

\begin{figure}
\begin{center}
\includegraphics[scale=0.6]{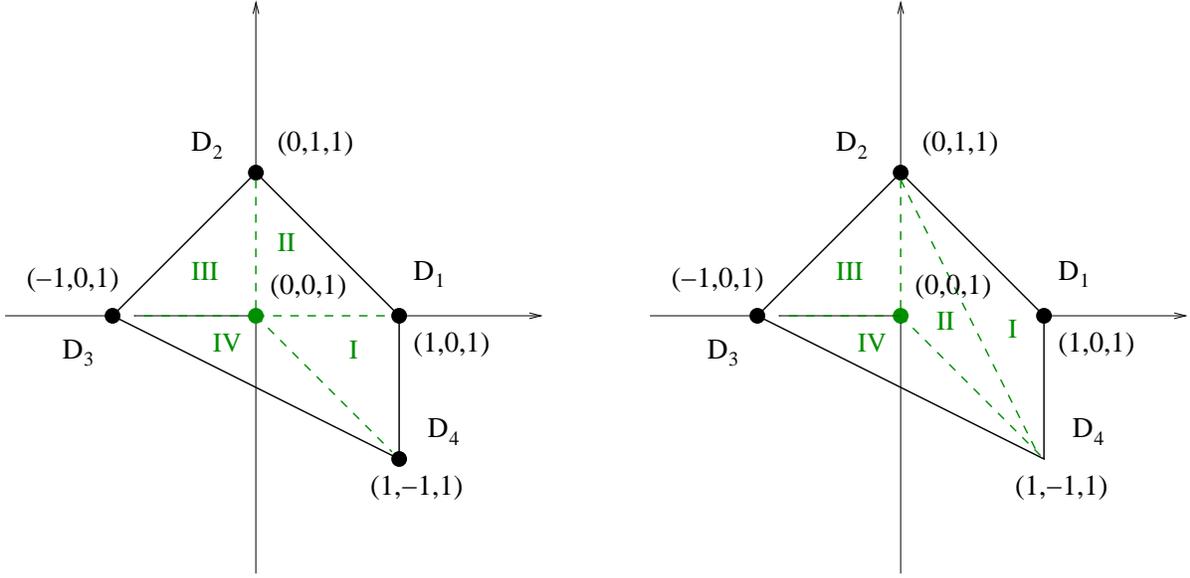} 
\caption{Two triangulation for the  toric diagram of $Y^{2,1}$. The internal point $(0,0,1)$ has been blown up. One line 
and four planes have been added to the original fan. 
There are four maximal cones and corresponding fixed points, denoted I,II,III and IV.}
\label{triang}
\end{center}
\end{figure}

Toric Calabi-Yau cones have a pretty standard resolution by triangulation
of the toric diagram, see Figure \ref{triang}. 
The fan of the
original variety $X$ consists of a single maximal cone, with a set of
edges, or one-dimensional cones, $\Sigma(1)$ whose generators $n_i$ are not
linearly independent in $\mathbb{Z}^3$. The 
resolutions of $X$ consist of all the possible subdivisions of the fan in
smaller three dimensional cones $\sigma_I$. The new variety $\tilde X$
is still a Calabi-Yau if all the minimal generators $n_i$ of the 
one-dimensional cones lie on a plane. This process looks like 
a triangulation of the toric diagram. If each three-dimensional cone is 
generated by linearly independent primitive vectors, the variety is smooth.
The smooth Calabi-Yau resolutions of $X$ thus consist of all the triangulation
of the toric diagram which cannot be further subdivided. 
Each three dimensional cone $\sigma_I$ is now a copy of $\mathbb{C}^3$ and
the smooth variety $\tilde X$ is obtained by gluing all these $\mathbb{C}^3$ 
according to the rules of the fan.
$T^3$ acts on each $\sigma_I$ in a simple way: the three
weights of the $T^3$ action on a copy of $\mathbb{C}^3$ are just given by
the primitive inward normal vectors $m_I^\lambda$ 
to the three faces of $\sigma_I$.
Notice that each $\sigma_I$ contains exactly one fixed point of $T^3$ (the origin in the copy of $\mathbb{C}^3$) with weights given by the vectors 
$m_I^\lambda$.

The line bundles on $\tilde X$ are given by $\tilde D=\sum_i c_i D_i$
where the index $i$ runs on the 
set of one-dimensional cones $\tilde\Sigma(1)$, which is typically bigger than
the original $\Sigma(1)$. Indeed, each integer internal point of the toric 
diagram gives rise in the resolution $\tilde X$ to a new divisor. The space of sections of $\tilde D$ are still determined by the 
integral points of the polytope
\begin{equation}\label{poly2}
\tilde P_D = \{ u \in M_{\mathbb{R}}| <u,n_i >\,  \ge\, - c_i\, , \,\,  \forall i\in \tilde 
\Sigma(1) \}
\end{equation} 
It is important for our purposes that each maximal cone $\sigma_I$ determines a integral point $m_I^0\in M$ as the
solution of this set of three equations:
\begin{equation}\label{charge}
< m_I^0, n_i > = - c_i, \,\,\, n_i\in \sigma_I,  
\end{equation}
In a smooth resolution $\tilde X$ this equation has always integer solution 
since the three generators $n_i$ of $\sigma_I$ are a basis for $\mathbb{Z}^3$.
As shown in \cite{fulton}, $m_I^0$ is the charge of the local equation for
the divisor $\tilde D$ in the local patch $\sigma_I$. It is therefore the
weight of the $T^3$ action on the fiber of ${\cal O}(D)$ over the fixed
point contained in $\sigma_I$.

The strategy for computing $Z_D$ is therefore the following. We smoothly
resolve $X$ and find a divisor $\tilde D= \sum_i c_i D_i$ by assigning
values $c_i$ to the new one-dimensional cones in $\tilde \Sigma(1)$
that satisfies the two conditions
\begin{itemize}
\item{It has the same space of sections, $H^0(\tilde X,{\cal O}(\tilde D)) = H^0(X,{\cal O}(D))$. Equivalently, the polytope $\tilde P_D$ has the same integer points of $P_D$.}
\item{It has vanishing higher cohomology
$H^{i}(\tilde X,{\cal O}(\tilde D))=0, i\ge 1$. As shown in
\cite{fulton} this is the case if there exist integer 
points $m_I^0\in M$ that satisfy the convexity condition
\footnote{The $m_I^0$s determine a continuous piecewise linear function $\psi_D$ on
the fan as follows: in each maximal cone $\sigma_I$ the function $\psi_D$
is given by $<m_I^0, v>$, $v\in \sigma_I$. As shown in \cite{fulton},
the higher dimensional cohomology vanishes, $H^i(\tilde X, {\cal O}(D))=0, \, i\ge 1$, whenever the function $\psi_D$ is upper convex.}
\begin{eqnarray}
< m_I^0, n_i > &=& - c_i, \,\,\, n_i\in \sigma_I \nonumber\\
< m_I^0, n_i > &\ge& - c_i, \,\,\, n_i\notin \sigma_I\label{convex}
\end{eqnarray}
}
\end{itemize}
There are many different smooth resolution of $X$, corresponding to
the possible complete triangulation of the toric diagram. It is shown
in the Appendix \ref{cicciop} that we can always find a compatible resolution $\tilde X$
and a minimal choice of $c_i$ that satisfy the two given conditions.

The function $Z_D$ is then given as
\begin{equation}
Z_D = \sum_{P_I} \frac{q^{m_I^0}}{\prod_{\lambda=1}^3 (1-q^{m_I^\lambda})}
\label{sumzd}
\end{equation}
where in the toric case for every fixed point $P_I$ there is a maximal cone $\sigma_I$, $m_I^\lambda$ are the three inward primitive normal vectors of 
$\sigma_I$ and $m_I^0$ are determined by equation (\ref{convex}). 
This formula can be conveniently generalized to the case where the fixed points are not isolated but there are curves or surfaces fixed by the torus action. 

We finish this Section with two comments. The first is a word of caution. Note
that if we change representative for a divisor in its equivalence class
($c_i\sim c_i+<M,n_i>$) the partition function $Z_D$ is not invariant, 
however it is just rescaled by a factor $q^M$.

The second comment concerns toric cones. For toric CY cones there is an
alternative way of computing the partition functions $Z_D$ by expanding
the homogeneous coordinate ring of the variety according to the decomposition
(\ref{dirs}). Since the homogeneous coordinate ring is freely generated by the $x_i$, 
its generating function is simply given by 
$$\frac{1}{\prod_{i=1}^d (1-x_i)}\, .$$
By expanding this function according to the grading given
by the $(\mathbb{C}^*)^{d-3}$ torus action we can extract all the $Z_D$.
This approach will be discussed in detail in a future publication \cite{ADAZ}. 

\subsection{Examples}
\subsubsection{The conifold}\label{conifoldexample}
The four primitive generators for the one dimensional cones of the conifold 
are $\{(0,0,1),(1,0,1),(1,1,1),(0,1,1)\}$ and we call the associated divisors $D_1,D_2,D_3$ and
$D_4$ respectively. They satisfy the equivalence relations $D_1\sim D_3\sim -D_2\sim -D_4$. There is only one baryonic symmetry under which
the four homogeneous coordinates transform as
\begin{equation} (x_1,x_2,x_3,x_4)\sim (x_1 \mu ,x_2/\mu,x_3 \mu ,x_4/\mu )\end{equation}
The conifold case is extremely simple in that the chiral fields of the 
dual gauge theory are in one-to-one correspondence with the homogeneous
coordinates: $(x_1,x_2,x_3,x_4)\sim (A_1,B_1,A_2,B_2)$. Recall that
the gauge theory is $SU(N)\times SU(N)$ with chiral fields $A_i$ and
$B_p$ transforming as $(N,\bar{N})$ and $(\bar{N},N)$ and as $(2,1)$ and $(1,2)$ under the enhanced $SU(2)^2$ global flavor symmetry.

\begin{figure}
\begin{center}
\includegraphics[scale=0.6]{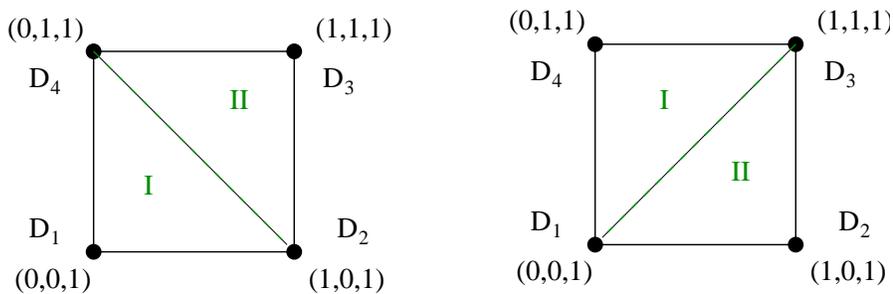} 
\caption{The two resolutions for the conifold. 
No internal points have been blown up. In each case, one line have been added to the original fan; there are two maximal cones and corresponding fixed points, denoted I,II.}
\label{conifold}
\end{center}
\end{figure}

The two possible resolutions for the conifold are presented in Figure \ref{conifold}. 
We first compute the partition function for the divisor $D_1$ using the
resolution on the left hand side of the figure. Regions $I$ and $II$ correspond
to the two maximal cones in the resolution and, therefore, to the  two fixed
points of the $T^3$ action. Denote also $q=(q_1,q_2,q_3)$. Using the prescriptions given above, we compute the three primitive inward 
normals to each cone  and the weight of
the $T^3$ action on the fiber. It is manifest that the
conditions required in equation (\ref{convex}) are satisfied.
\begin{eqnarray}
{\rm Region\, I}\qquad &m_I^\lambda = \{ ((1,0,0),(0,1,0),(-1,-1,1)\}\qquad & m_I^0=(1,1,-1)\nonumber\\
{\rm Region\, II}\qquad & m_{II}^\lambda = \{
(0,-1,1),(-1,0,1),(1,1,-1)\}\qquad & m_{II}^0=(0,0,0)\nonumber
\end{eqnarray}
\vskip -0.5truecm
\begin{equation}
 Z_{D_1} = \frac{q_1(q_2-q_3)+q_3-q_2 q_3}{(1-q_1)(1-q_2)(1-q_3/q_1)(1-q_3/q_2)q_3}\label{conD1}
\end{equation}
For simplicity, let us expand $Z_{D_1}$ along the direction of  the Reeb vector $(3/2,3/2,3)$ by putting $q_1=q_2=q,q_3=q^2$. This corresponds to count
mesonic excitations 
according to their R-charge, forgetting about the two $U(1)^2$ flavor indices. 
\begin{equation}
Z_{D_1}=\frac{2}{(1-q)^3}=\sum_{n=0}^\infty (n+1)(n+2) q^n = 2 + 6 q + 12 q^2 + ... 
\end{equation}
This counting perfectly matches the list of operators in the gauge 
theory. In the sector of Hilbert space with  baryonic charge $+1$ we find
the operators (\ref{AAA})
\begin{equation}
A_i\, , \qquad A_i B_p A_j\, , \qquad A_i B_p A_j B_q A_k\, , \qquad ....
\end{equation}
The F-term equations $A_i B_p A_j=A_j B_p A_i$, $B_p A_i B_q = B_q A_i B_p$
guarantee that the $SU(2)\times SU(2)$ indices are totally symmetric. The
generic operator is then of the form $A(BA)^n$ transforming in the 
$(n+2,n+1)$ representation of $SU(2)\times SU(2)$ thus 
exactly matching the $q^n$ term in $Z_{D_1}$.  
The R-charge of the operators in $Z_{D_1}$ 
is accounted by the exponent of $q$ by adding the factor $q^{\sum c_i R_i}=q^{1/2}$ which is common to all the operators in this sector (cfr. equation (\ref{erre})). The result perfectly matches with the operators  $A(BA)^n$ since the 
exact R-charge of $A_i$ and $B_i$ is $1/2$.
We could easily include the $SU(2)^2$ charges in this counting.

Analogously, we obtain for $Z_{D_3}$
\begin{equation}
 Z_{D_3} = \frac{q_1(q_2-q_3)+q_3-q_2 q_3}{(1-q_1)(1-q_2)(q_1-q_3)(q_2-q_3)}
= q_3 Z_{D_1}/(q_2 q_1)\label{conD3}
\end{equation}
Since $D_1\sim D_3$ the polytope $P_{D_3}$ is obtained by $P_{D_1}$ by a
translation and the the two partition functions $Z_{D_1}$ and $Z_{D_3}$ are
proportional. Finally, the partition functions for $D_2$ and $D_4$ are obtained by choosing the resolution in the right hand side of figure \ref{conifold}, for which is possible to satisfy the convexity condition (\ref{convex})
\begin{eqnarray}
Z_{D_2} &=& \frac{q_2 (q_1 + q_2 - q_1 q_2 -q_3)}{(1 - q_1)(1 - q_2)(q_1 - q_3)(q_2 -q_3)}\nonumber\\
Z_{D_4} &=& \frac{q_1 (q_1 + q_2 - q_1 q_2 -q_3)}{(1 - q_1)(1 - q_2)(q_1 - q_3)(q_2 -q_3)} = q_1 Z_{D_2} /q_2
\end{eqnarray}

\subsubsection{Other examples: $Y^{p,q}$, delPezzo and $L^{p,q,r}$}
In this Section we give other examples of partition functions $Z_D$ considering
the $Y^{p,q}$, the delPezzo  and $L^{p,q,r}$ manifolds. 

The $Y^{p,q}$ toric diagram has four vertices and one baryonic charge.
The dual gauge theory has an $SU(2)\times U(1)$ flavor symmetry.
We consider the simplest example, $Y^{2,1}$. 
The fan for $Y^{2,1}$ has four primitive generators 
$\{(1,0,1),(0,1,1),(-1,0,1),(1,-1,1)\}$. The equivalence relations among
divisors give $D_2\sim D_4\sim -2 D_3$ and $D_1=3 D_3$ and the corresponding
homogeneous coordinates scale as
\begin{equation}
(x_1,x_2,x_3,x_4)\sim (x_1 \mu^3, x_2/\mu^2,x_3 \mu, x_4/\mu^2)
\end{equation}
under the baryonic symmetry. 

There are two different completely smooth resolutions that are presented in
Figure \ref{triang}. The toric diagram has one internal 
point; the corresponding four cycle is blown up in each smooth resolution
of the cone and introduces a new divisor $D_5$. 
In each resolution there are four fixed points for
the action of $T^3$. 

To compute the partition functions we need to chose a resolution
and the number $c_5$ that satisfy the convexity condition (\ref{convex}). 
The partition function for $Z_{D_3}$ can be
computed by using the resolution on the left hand side in the figure and
the number $c_5=0$.
\begin{eqnarray}
{\rm Region\, I}\qquad &m_I^\lambda = \{ ((-1,0,1),(0,-1,0),(1,1,0)\}\qquad & m_I^0=(0,0,0)\nonumber\\
{\rm Region\, II}\qquad &m_{II}^\lambda = \{
((-1,-1,1),(0,1,0),(1,0,0)\}\qquad & m_{II}^0=(0,0,0)\nonumber\\
{\rm Region\, III}\qquad & m_{III}^\lambda = \{
(1,-1,1),(0,1,0),(-1,0,0)\}\qquad & m_{III}^0=(1,0,0)\nonumber\\
{\rm Region\, IV}\qquad & m_{IV}^\lambda = \{
(1,2,1),(0,-1,0),(-1,-1,0)\}\qquad & m_{IV}^0=(1,1,0)\nonumber
\end{eqnarray}
\vskip -0.5truecm
\begin{eqnarray}
Z_{D_3} = \frac{-q_3^2+q_2^2(q_1^2q_3-q_3^2+q_1(1+q_3-q_3^2))-q_2(-1+q_3+q_3^2-q_1 -q_1 q_3+q_1q_3^2)}{(1- q_3/(q_1 q_2))(1- q_3/q_1)(q_2-q_1 q_3) (1 - q_1 q_2^2 q_3)} \nonumber
\end{eqnarray} 
$Z_{D_3}$ can be expanded using the 
geometric series by setting $q_3=q q_1 q_2$.
It is immediate to verify that the first terms in the expansion 
$Z_{D_3}=1+q_1 +q_1q_2+ q(q_2+1+q_1+...)+...$ exactly match the list of
field theory operators given in Section 5 (cfr Table 1).

The partition functions for the other three elementary divisors can be computed
in a similar way. In order to satisfy the
 convexity condition we use the resolution on the left of figure \ref{triang}
for $D_2$ and $D_4$ and the resolution on the right for $D_1$. In all cases we can safely put $c_5=0$.
\begin{eqnarray}
&&Z_{D_1} = \frac{q_1^2q_2^2+q_2(1+q_1^2(1+(1+q_1)(q_2+q_2^2)))q_3-q_1(1+(1+q_1)(q_2+q_2^2+q_2^3))q_3^2}{(q_1 q_2- q_3)(q_1- q_3)(q_2-q_1 q_3) (1 - q_1 q_2^2 q_3)}\nonumber\\
 && Z_{D_2} = \frac{q_1^2q_2^2 q_3-q_1^2q_2^2q_3^2-q_3(q_2+(1+q_2+q_2^2)q_3)+q_1(1+q_2)(q_3+q_2+q_2^2q_3-q_2q_3^2)}{(q_1 q_2- q_3)(1- q_3/q_1)(q_2-q_1 q_3) (1 - q_1 q_2^2 q_3)}  \nonumber\\
&&Z_{D_4} = q_2 Z_{D_2}\nonumber
\end{eqnarray}
The proportionality of $Z_{D_4}$ and $Z_{D_2}$ follows from the equivalence
$D_2\sim D_4$.

\begin{figure}
\begin{center}
\includegraphics[scale=0.4]{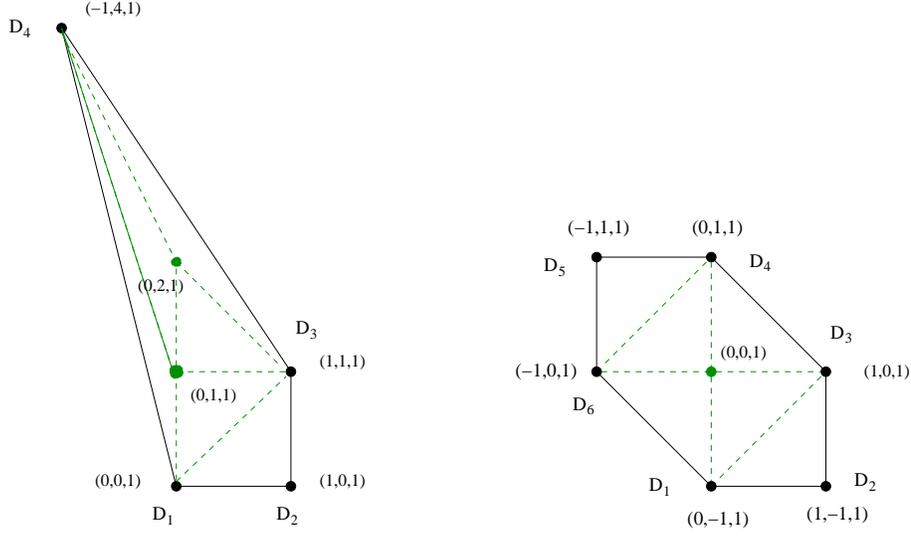} 
\caption{$L^{1,5,2}$ and $dP_3$. On the left: a compatible resolution for the $D_4$ partition function of $L^{1,5,2}$, with six fixed points. On the right: a compatible resolution for the $D_2$ partition function for $dP_3$, with six fixed
points. In both cases, one can safely choose the $c_i$ of the extra divisors
equal to zero.}
\label{LdP}
\end{center}
\end{figure}

Similarly, one can compute the partition functions for the other $Y^{p,q}$ 
manifolds and, more generally, for the $L^{p,q,r}$ manifolds which 
correspond to the most general toric diagram with four external points.
The flavor symmetry for $L^{p,q,r}$ is $U(1)^2$ and, for smooth
manifolds, there is exactly one baryonic symmetry. The number of internal 
points increases with $p,q,r$ thus making computations more involved.
As an example, we present the partition function for the $D_4$ divisor in 
$L^{1,5,2}$. We refer to Figure \ref{LdP}
for notations and the choice of a compatible resolution. 
\begin{eqnarray}
Z_{D_4} &=& \frac{P(q_1,q_2,q_3)}{(1-q_2)(1-q_1^4 q_2)(q_1-q_3)(q_1^3 q_2^2-q_3^5)}\nonumber\\
P(q_1,q_2,q_3) &=& q_1q_2 (q_1^6q_2^2-q_3^5-q_1q_3^5-q_1^2 q_3^2 (-1+q_2-q_3+q_2q_3+q_3^3)\nonumber\\
& + &q_1^5 q_2 (q_3(1+q_3)-q_2(-1+q_3+q_3^2))
+ q_1^4 (-q_2^2 q_3 +q_3^4 +q_2(1+q_3-q_3^4))\nonumber\\
&+&q_1^3 (q_3^3+q_3^4-q_3^5-q_2(-1+q_3^3+q_3^4)))
\end{eqnarray}
Finally, we give an example based on the $dP_3$ manifolds, whose toric diagram
has six external points and thus three different baryonic symmetries. Once
again we refer to Figure \ref{LdP} for notations.
\begin{eqnarray}
&&Z_{D_2} = \frac{P(q_1,q_2,q_3)}{(q_1-q_3)(1-q_3/q_2)(q_3-q_1q_2)(1-q_1q_3)(1-q_2q_3)(1-q_1q_2q_3)}\nonumber\\
&&P(q_1,q_2,q_3) = (q_2+q_1(1+2q_2+q_1(1+q_2)(1+q_2(1+q_1+q_2))))q_3^2\nonumber\\
&&-(1+q_1((1+q_2)^2+q_1(1+q_2)(1+2q_2)+q_1^2q_2(1+q_2+q_2^2)))q_3^3\nonumber\\
&&+q_1(1+(1+q_1)q_2(1+q_1q_2))q_3^4-q_1^2 q_2-q_1q_2(q_1+q_2)q_3\nonumber
\end{eqnarray}
Following this procedure one is able to compute the partition function $Z_D$ for every divisor of a generic $CY$ conical toric singularity.

\subsection{The partition function for BPS baryonic operators}
The BPS baryonic states in a sector of the Hilbert space associated
with the divisor $D$ are obtained from the $h_m$ by considering
the N-fold symmetrized combinations $|h_{m_1},....,h_{m_N}>$. The 
partition function $Z_{D,N}$ for BPS baryon is obtained from $Z_D$ by
solving a combinatorial problem \cite{Kinney:2005ej,Benvenuti:2006qr}. 

Given $Z_D$ as sum of integer points in the polytope $P_D$
\begin{equation}
Z_D(q) = \sum_{m\in P_D \cap M} q^m
\end{equation}
the generating function $G_D(p,q)$ for symmetrized products of elements in $Z_D$ is
given by
\begin{equation}
G_D(p,q)=\prod_{m\in P_D \cap M} \frac{1}{1-p q^m} =\sum_{N=0}^\infty p^N Z_{D,N}(q)
\label{multi}\end{equation}
This formula is easy to understand: if we expand $(1-pq^m)^{-1}$ in 
geometric series, the coefficient of the term $p^k$ is given by 
all possible products of $k$ elements in $P_D$, and this is clearly a $k$-symmetric product.\\
It is easy to derive the following relation between $Z_D(q)$ and $G_D(p,q)$
\begin{equation}\label{expG}
G_D(p,q)=e^{\sum_{k=1}^\infty \frac{p^k}{k} Z_D(q^k)}
\end{equation}

In the case we have computed $Z_D(q)$ in terms of  the fixed point data of a compatible resolution as in equation  (\ref{sumzd}) 
$$Z_D(q)=\sum_I \frac{q^{m_I^0}}{\prod_{\lambda=1}^3(1-q^{m_I^\lambda})}=\sum_I\sum_{s_I^1,s_I^2,s_I^3} q^{m_I^0}q^{\sum_{\lambda=1}^3 s_I^\lambda m_I^\lambda}$$
formula (\ref{expG}) allows, 
with few algebraic manipulation, to write the generating
function  as follows
\begin{equation}
G_D(p,q)=\prod_{P_I}\prod_{s_I^\lambda=0}^\infty \frac{1}{1- p q^{m_I^0} q^{\sum_{\lambda=1}^3 s_I^\lambda m_I^\lambda}}
\end{equation}

 We are eventually interested
in the case of BPS baryonic operators associated with the symmetrized
elements $|h_{m_1},....,h_{m_N}>$, and thus to the $N$-fold symmetric partition function:
\begin{equation}\label{sympar}
Z_{D,N}(q)\equiv \frac{1}{N!}\frac{\partial ^N G_D(p,q)}{\partial p^N } \Big| _{p=0}
\end{equation}
Thanks to $G_D(0,q)=1$ (see eq. (\ref{expG})) we can easily write $Z_{D,N}$ in function of $Z_D$.
For example we have:
$$Z_{D,1}(q)= Z_D(q)$$
$$Z_{D,2}(q) = \frac{1}{2} (Z_D(q^2)+Z_D^2(q))$$
$$Z_{D,3}(q) = \frac{1}{6} (2 Z_D(q^3)+ 3 Z_D(q^2) Z_D(q) + Z_D^3(q))$$
Once we know $Z_D$ for a particular baryonic sector of the $BPS$ Hilbert space it is easy to write down the complete partition function $Z_{D,N}$.\\
\\

\section{Volumes of divisors}
One of the predictions of the AdS/CFT correspondence for the
background $AdS_5\times H$ is that the volume of $H$ is related
to the central charge $a$ of the CFT, and the volumes of the three cycles
wrapped by the D3-branes are  related to the R-charges of the corresponding
baryonic operators \cite{Gubser:1998fp,Gubser:1998vd}. We already used this information in formula
(\ref{errevol}). To many purposes, it is useful to consider the volumes
as functions of the Reeb Vector $b$.
Recall that each K\"ahler metric on the cone, or equivalently a Sasakian structure on the base $H$,  determines a Reeb vector $b=(b_1,b_2,b_3)$ and
that the knowledge of $b$ is sufficient to compute all volumes in $H$ \cite{MSY}. 
Denote with $\rm{Vol}_H(b)$ the volume of the base of a K\"ahler cone with Reeb vector $b$.
The Calabi-Yau condition $c_1(X)$ requires $b_3=3$ \cite{MSY}. As shown in \cite{MSY,MSY2}, the Reeb vector $\bar b$ 
associated with the Calabi-Yau metric can be obtained by minimizing the function $\rm{Vol}_H(b)$ with respect to $b=(b_1,b_2,3)$. This volume minimization is the geometrical 
counterpart of a-maximization in field theory \cite{intriligator}; the equivalence of a-maximization and volume minimization has been explicitly proven for all toric cones in \cite{aZequiv} and for a class a non toric cones in \cite{Butti:2006nk}. For each Reeb vector $b=(b_1,b_2,b_3)$ we can also define the
volume of the three cycle obtained by restricting a divisor $D$ to the base,
$\rm{Vol}_{D}(b)$. We can related the value $\rm{Vol}_{D}(\bar b)$ at the 
minimum to the exact R-charge of the lowest dimension baryonic operator associated with the divisor $D$ \cite{Gubser:1998fp,Gubser:1998vd} as in formula (\ref{errevol}).

All the geometrical information about volumes can be
extracted from the partition functions.
The relation between the character $Z_{\cal{O}}(q)$ 
for holomorphic functions on $C(H)$ and the volume of $H$ was suggested
in \cite{Bergman:2001qi} and proved for all K\"ahler cones in \cite{MSY2}. If we
define $q=(e^{-b_1 t},e^{-b_2 t},e^{-b_3 t})$, we have \cite{Bergman:2001qi,MSY2}
\begin{equation}
\rm{Vol}_H (b) =\pi^3 \lim_{t\rightarrow 0} t^3 Z_{\cal{O}} (e^{-b t})\label{vol}
\end{equation}
This formula can be interpreted as follows: the partition function $Z_{\cal{O}}(q)$ has a pole for $q\rightarrow 1$, and the order of the pole - three - reflects the complex dimension of $C(H)$ while
the coefficient is related to the volume of $H$. 

Here we propose that, similarly, the partition functions $Z_D$ contain the information about the three-cycle volumes $\rm{Vol}_{D}(b)$. Indeed we suggest that,
for small $t$ \footnote{And a convenient choice of $D$ in its equivalence class.},
\begin{equation}
\frac{Z_D(e^{-b t})}{Z_{\cal{O}}(e^{-b t})} \sim 1 + t \frac{\pi  \rm{Vol}_D(b)}{2 \rm{Vol}_H(b)}+...\label{voldiv}
\end{equation}
Notice that the leading behaviour for all partition functions $Z_D$ is the same and proportional to the volume of $H$; for $q\rightarrow 1$ the main contribution comes from states with arbitrarily large dimension and it seems that
states factorized in a minimal determinant times gravitons dominate the
partition function. The proportionality to $\rm{Vol}_H$ is then expected
by the analogy with giant gravitons probing the volume of $H$. The subleading
term of order $1/t^2$ in $Z_D$ then contains information about the dimension 
two complex divisors. 
Physically it is easy to understand that $Z_D$ contains the information about the volumes of the divisors. 
We can think at $Z_D$ as a semiclassical parametrization of the holomorphic non-trivial surfaces in $X$, 
with a particular set of charges related to $D$; while $Z_{\mathcal{O}}$ parametrizes the set of trivial 
surfaces in $X$. Thinking in this way it is clear that both know about the volume of the compact space, 
but only $Z_D$ has information on the volumes of the non-trivial three cycles. 

For divisors $D$ associated with elementary fields we can rewrite 
equation (\ref{voldiv}) in a simple and suggestive way in terms 
of the R-charge, or dimension,  of the elementary field (see equation (\ref{errevol}))
\begin{equation}
\frac{Z_D(e^{-b t})}{Z_{\cal{O}}(e^{-b t})} \sim 1+ t \frac{3 R_D(b)}{2}+...=1+ t \Delta(b)+...
\end{equation}
 
 As a check of formula (\ref{voldiv}), we can expand the partition functions
for the conifold computed in the previous Section
\begin{equation}
\frac{Z_{D_i}}{Z_{\cal{O}}}\sim(1+\frac{(b_1-b_3)(b_2-b_3)t}{b_3},1+\frac{(b_1 b_3-b_1 b_2)t}{b_3},1+\frac{b_1 b_2 t}{b_3},1+\frac{(b_2 b_3 -b_1 b_2)t}{b_3})
\nonumber
\end{equation}
 and compare it with the formulae in \cite{MSY}
\begin{equation}
\rm{Vol}_{D_i}(b)=\frac{2\pi^2 \det \{n_{i-1},n_i,n_{i+1}\}}{\det\{b,n_{i-1},n_i\} \det \{b,n_i,n_{i+1}\}}\, , \qquad \rm{Vol}_H(b)=\frac{\pi}{6}\sum_{i=1}^d \rm{Vol}_{D_i}(b)\label{tor}
\end{equation}
One can perform similar checks for $Y^{2,1}$ and the other cases considered
in the previous section, with perfect agreement. A sketch of a general 
proof for formula (\ref{voldiv}) is given in the Appendix \ref{Proof}.

We would like to notice that, by expanding equation (\ref{sumzd}) for $q=e^{-b t}\rightarrow 1$ and comparing with formula (\ref{voldiv}), we are able to write
a simple formula for the volumes of divisors in terms of the fixed point
data of a compatible resolution
 \begin{equation}
 \rm{Vol}_D(b)=2 \pi^2 \sum_{P_I} \frac{(-m_I^0,b)}{\prod_{\lambda = 1}^3 (m_I^\lambda, b)}
\label{volsum}
\end{equation}
This formula can be conveniently generalized to the case where the fixed points are not isolated but there are curves or surfaces fixed by the torus action.

The previous formula is not specific to toric varieties. It can be used  whenever we are able to resolve the cone $C(H)$ and to reduce the computation of $Z_D$ to a sum over isolated fixed points (and it can be generalized to the case 
where there are fixed submanifolds). As such, it 
applies also to non toric cones. The relation between
volumes and characters may give a way for computing volumes of
divisors in the general non toric case, where explicit formulae like 
(\ref{tor}) are not known. 

\section{Conclusion and Outlook}

In this paper we proposed a general procedure to construct partition functions counting both 
baryonic and non baryonic $BPS$ operators of a field theory dual to a toric geometry. 
We also explained how one can extract the volumes of the three cycles from the partition functions. 
It would be interesting to understand better the counting of multiplicity, and to find a way 
to write down a complete partition function for the $BPS$ gauge invariant scalar operators \cite{ADAZ}. 

Our computation is done on the supergravity side, and it is therefore valid
at strong coupling. Similarly to 
the partition function for BPS mesonic operators \cite{Kinney:2005ej,Biswas:2006tj,Benvenuti:2006qr}, 
we expect to be able to extrapolate the result to finite value for the coupling. 

It would be also interesting to understand better the non toric case. Most
of the discussions in this paper apply to this case as well. The classical 
configurations of BPS D3 branes wrapping a divisor $D$ are still 
parameterized by the generic section of $H^0(X,{\cal O}(D))$ and Beasley's 
prescription for constructing the BPS Hilbert space is unchanged. The partition
function $Z_D(q)$ can be still defined, with the only difference that 
$q\in T^k$ with $k$ strictly less than three. $Z_D(q)$ can be still
computed by using the index theorem as explained in Section \ref{counting}
and the relation between $Z_D(q)$ and the three cycles volumes should be
still valid. In particular, when $X$ has a completely smooth resolution
with only isolated fixed points for the action of $T^k$, formulae
 (\ref{sumzd}) and (\ref{volsum}) should allow to compute the partition functions and 
the volume. What is missing in the non-toric case is an analogous of the
homogeneous coordinates, the polytopes and the existence of a canonical
smooth resolution of the cone. But this seems to be just a technical problem.

\vspace{2em} 
\noindent {\Large{\bf Acknowledgments}}

\vspace{0.5em}

We would like to thank Amihay Hanany  and Alessandro Tomasiello for useful discussions.
D.F. thanks Jose D. Edelstein, Giuseppe Milanesi and Marco Pirrone 
for useful discussions. 
This work is supported in part by INFN and MURST under  
contract 2005-024045-004  and 2005-023102 and by 
the European Community's Human Potential Program
MRTN-CT-2004-005104. 
  
\appendix
\section{Localization formulae for the volumes of divisors}\label{Proof}
Relation (\ref{voldiv}) can be easily proven in the case where the Reeb
action is regular, by adapting an argument in \cite{Bergman:2001qi,Herzog:2003wt}, as refined
in \cite{MSY2}. For a regular action, $H$ is a $U(1)$ principal bundle
over a K\"ahler manifold $V$ and $X$ can be written as a line bundle 
$L\rightarrow V$. We can blow up $V$ and apply equivariant Riemann-Roch
to the resulting manifold. Since the Reeb vector acts on the fibers of
$L$, its fixed locus is the entire $V$ with weight $1$. We thus
obtain \footnote{The multiplicative ambiguity in $Z_D$ is reflected in an analogous ambiguity in $Ch(D)$.}
\begin{equation}
Z_D(q)=\int_V \frac{{\rm Todd}(V) Ch(D)}{ 1-q e^{-c_1(L)} }
\label{pio}\end{equation}
Put $q=e^{-b t}$ in this formula. 
The denominator must be expanded
in a formal power series of forms before taking the limit $t\rightarrow 0$
$$\frac{1}{1-q e^{-c_1(L)}}=\frac{1}{1-e^{-b t}}-\frac{e^{-b t}}{(1-e^{-b t})^2}c_1(L)+\frac{e^{- b t}+e^{-2 b t}}{2(1-e^{-b t})^3}c_1(L)^2$$
Since the integral over $V$ selects forms of degree four we obtain
$$Z_D(q)=\frac{1}{(b t)^3}\int_V c_1(L)^2 -\frac{1}{(b t)^2}\int_V c_1(L)\wedge {\rm Todd}(V)\wedge Ch(D)\Big |_{{\rm degree}\, 4}+...$$   
The only information we need about the Todd class is that ${\rm Todd}(V)=1+...$. We thus obtain
$$\frac{Z_D(q)}{Z_{\cal O}}=1 - b\, t\, \frac{\int_V c_1(L)\wedge c_1(D)}{\int_V c_1(L)\wedge c_1(L)}+...$$

The volumes of $H$ and of the three cycle
$D\cap H$, which are $U(1)$ fibrations over $V$ and $D\cap V$, are proportional to
\begin{eqnarray}
{\rm Vol}_D (b)&\sim&\int_D \omega_V=\int_V \omega_V\wedge c_1(D)\nonumber\\
{\rm Vol}_H (b)&\sim&\int_V \frac{\omega_V^2}{2}
\end{eqnarray}
Considering that the first Chern class of $L$
is proportional to the K\"ahler form $\omega_V$ on $V$ 
\footnote{We use the normalizations
of \cite{MSY2}: $c_1(L)=-b c_1(V)/3$ and $\omega_V=\pi c_1(V)/3$. These
formulae are valid also for a quasi regular action. The length of the $U(1)$ fiber is $2\pi/b$.}, we finally obtain
$$\frac{Z_D(q)}{Z_{\cal O}}= 1 + t \frac{\pi  \rm{Vol}_D(b)}{2 \rm{Vol}_H(b)}+...$$

Formula (\ref{voldiv}) could be proven for a generic Sasaki-Einstein
by generalizing arguments in \cite{MSY2}. We only suggest a possible proof,
leaving to experts the subtle task of filling mathematical details.
As argued in \cite{MSY2}, it is enough to prove (\ref{voldiv}) for quasi
regular actions; since a rational $b\in T^k$ defines a Sasaki structure
on $H$ with quasi regular Reeb action, formula (\ref{voldiv}) would be true
for all rational $b$ and, therefore, for continuity, for all $b$. For
a quasi regular action, $L\rightarrow V$ is an orbifold and we should use
the Kawasaki-Riemann-Roch formula  \cite{kawasaki} which have extra contributions with
respect to (\ref{pio}). However, for isolated orbifold singularities, the
extra contributions are characteristic classes integrated over points; 
these contribute to $Z_D(q)$ only at order $1/t$ and should be 
irrelevant for our purposes. 

It would be interesting to fully understand formula (\ref{voldiv}) in terms 
of localization. It seems that some localization theorem is 
at work here. Considering that the  action of the Reeb field
$$\xi=\sum_{i=1}^k b_i \frac{\partial}{\partial\phi_i}$$
(we have chosen $k$ angular coordinates for the torus $T^k$ action, 
$k=1,2,3$) is hamiltonian, we can define the equivariantly closed form
$\omega^{\xi}=\omega - H$ starting from the k\"ahler class $\omega$ \footnote{ Given a vector field $\xi$ the equivariant derivative $d_{\xi}$ of a form $\alpha$ is $d_{\xi}\alpha = d \alpha + i_{\xi}\alpha$; $\omega ^{\xi}$ is clearly equivariantly closed, because $\omega$ is closed and $H$ is the Hamiltonian of the Reeb vector field ($i_{\xi}\omega=dH$).}. 
As shown in \cite{MSY2}, the hamiltonian for the Reeb action is $H=r^2/2$. Analogously we can define the equivariant first Chern class $c_1^{\xi}(D)$ associated with the divisor $D$.  
It is interesting to notice that the  expression (\ref{volsum}) for the volumes can be written as the integral of equivariantly characteristic classes,
\begin{equation}
\frac{1}{2}\int_{X}  e^{\omega^{\xi}}\wedge c_1^{\xi}(D) =  2 \pi^2
\sum_{\sigma_I} \frac{(-m_I^0,b)}{\prod_{\lambda = 1}^3 (m_I^\lambda, b)}
\label{volsum2}
\end{equation}

This equality can be proven as follows.
Suppose that we have found a smooth resolution $\tilde X$ of the cone $X$
and a divisor $\tilde D$ that smoothly approach $D$ in the singular limit.
We may then compute the previous integral for $\tilde X$ and $\tilde D$
and take afterwards the limit $\tilde X\rightarrow X$.
Integrals of equivariantly closed forms, like (\ref{volsum2}), can be
computed by using localization theorems. Indeed given an equivariantly
closed form $\alpha$ and an action along a direction in $T^k$ 
($k=1,2,3$ ) with only isolated fixed points, it can be shown that
\begin{equation}
\int \alpha = (2\pi)^3 \sum_{P_I} \frac{\alpha|_{P_I}}{\prod_{\lambda=1}^3
  (m_I^\lambda,b)} 
 \end{equation}
where $m_I^\lambda$ are, as usual, the weights of the action of $\xi$ on the
tangent space at $P_I$. The integral over a point $P_I$ takes contribution
only from the forms with zero degree in the equivariant forms,   
\begin{eqnarray}
&&\omega^{\xi}\rightarrow -H(P_I)\nonumber\\
&&c_1^{\xi}(D)\rightarrow -\frac{(m_I^0,b)}{2\pi}
\end{eqnarray}
where $m_I^0$ is the weight of the action on the line bundle fiber over $P_I$
\footnote{The second formula follows from the standard replacement $c_1\rightarrow c_1-w/(2\pi)$, with $w$ the weight for the group action, 
for line bundles over fixed submanifolds.}.
Taking into account that in the singular limit all the $P_I$s collapse
to the apex of the cone where $H = 0$, we finally obtain 
formula (\ref{volsum2}).
 
On the other hand, also the volume of the base $D\cap H$  can be written
as an integral
\begin{equation}
\rm{Vol_D}(b)=\frac{1}{2}\int_D e^{-r^2/2} \, \frac{\omega^2}{2} =  \frac{1}{2}\int_D e^{\omega^{\xi}} 
\end{equation}
Comparing this equation with the expression (\ref{volsum2}) for the volumes
we find the suggestive equality
\begin{equation}
\int_D e^{\omega^{\xi}}=\int_{X}  e^{\omega^{\xi}}\wedge c_1^{\xi}(D)\label{int}\, .
\label{proofa}\end{equation} 
concerning $T$ invariant divisors $D$.
Our general expression for the volumes (\ref{volsum}) (which, in case there
exist smooth resolutions for $X$ with isolated fixed points, is completely
equivalent to the general formula (\ref{voldiv})) would be proved if we were
able to prove equation (\ref{proofa}).
The relation between equivariant cohomology and homology seems to
be not completely understood (to us at least), and we do not know under what condition the equation (\ref{proofa}) is valid, even if this is probably well known to
experts.
A better understanding of this formula could give a simple
alternative proof for (\ref{voldiv}).

\section{Convexity condition and integer counting}\label{cicciop}
In this Appendix we give an alternative explanation of the formula (\ref{sumzd})
for computing the partition function $Z_D$ defined in (\ref{character}) and we
explain why there always exists a suitable resolution of $X=C(H)$ and a
suitable choice of weights $\tilde{c}_i$ for equation (\ref{sumzd}). In
Section \ref{riemannroch} we explained how to derive this formula from the equivariant
Riemann-Roch theorem. However since $Z_D$ counts the holomorphic sections of
$\mathcal{O}(D)$ and since we know that, for toric singularities, these
sections are
in one to one correspondence with the integer points inside the polytope
$P_D$, we can simply look at this problem as that of counting integer points
inside a polytope, with the weights $q=(q_1,q_2,q_3)$ being associated with
the three cartesian coordinates: $Z_D=\sum_{m \in P_D \cap M } q^m$. This simple point
of view allows to have a direct understanding of the counting problem. 

\begin{figure}
\begin{center}
\includegraphics[scale=0.45]{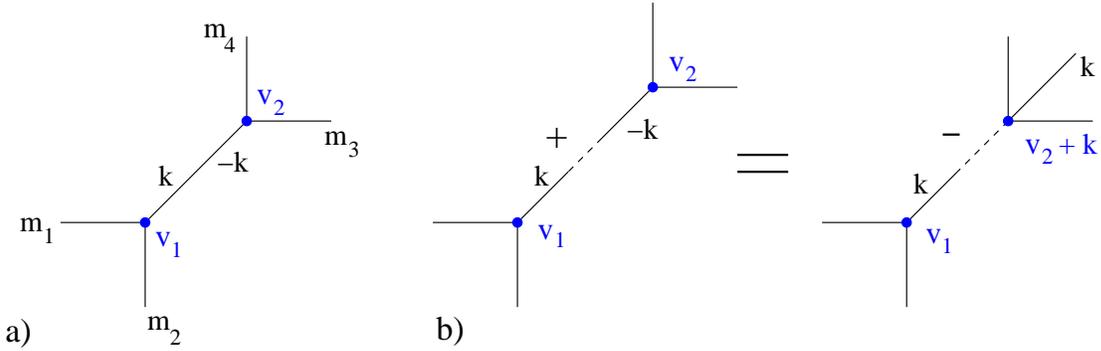} 
\caption{a) The (p,q) web for the conifold resolution corresponding to the first
diagram in Figure \ref{conifold}. b) A pictorial description of equation
(\ref{cancel}).}
\label{pqweb}
\end{center}
\end{figure}

To be concrete consider for instance the case of the conifold; let us write
the character that counts integer points inside the dual fan, which is
equivalent to counting holomorphic functions, or sections of the trivial
bundle $\mathcal{O}$. 
The dual fan is generated by the four vectors:
\begin{equation}
m_1=(1,0,0) \quad m_2=(0,1,0) \quad m_3=(-1,0,1) \quad m_4=(0,-1,1)
\end{equation}
all attached at the origin $n=(0,0,0)$.
To compute the character we will use for instance the first resolution in Figure
\ref{conifold}, whose corresponding (p,q) web is drawn in Figure \ref{pqweb} a). Let us
define the vector $k=(-1,-1,1)$. We split the (p,q) web into two subwebs
corresponding to region I and II respectively of the resolution in Figure
\ref{conifold}: 
\begin{equation}
\begin{array}{l@{\qquad}l@{\qquad \qquad}l}
\textrm{Region I} & \{ m_1, m_2, k \} & v_1=(0,0,0) \\
\textrm{Region II} & \{ m_3, m_4, -k \} & v_2=(0,0,0)
\end{array}
\end{equation}   
We denote with $v_1$ ($v_2$) the integer point to which the three vectors of
region I (II) are attached. In this case $v_1=v_2=(0,0,0)$.
Since we have completely resolved the conifold, according to \cite{MSY2}, the
character is:
\begin{equation}
Z_{\mathcal{O}}=\frac{1}{(1-q^{m_1})(1-q^{m_2})(1-q^{k})} +
\frac{1}{(1-q^{m_3})(1-q^{m_4})(1-q^{-k})}
\label{zetac}
\end{equation}
where as usual $q^h \equiv q_1^{h_1} q_2^{h_2} q_3^{h_3}$. It is simple to
give an interpretation of this formula in terms of counting of integer
points. Let us for instance expand the first term, associated with Region I,
in equation (\ref{zetac}). We get
\begin{equation}
\left( \sum_{i=0}^{\infty} q^{i\, m_1} \right)
\left( \sum_{j=0}^{\infty} q^{j\, m_2} \right)
\left( \sum_{h=0}^{\infty} q^{h\, k} \right)=
\sum_{i,j,h \geq 0} q^{i \, m_1 + j \, m_2 + h \, k}
\label{exp}
\end{equation} 
and this is just the partition function that counts integer points inside the
cone generated by the vectors $\{ m_1, m_2, k \}$ attached to the origin $v_1=(0,0,0)$.
In fact each integer point inside this cone can be written in a unique way as
a linear combination of  $\{ m_1, m_2, k \}$ with positive integers due to the
fact that  $ \textrm{det} ( m_1, m_2, k) = 1 $. This is equivalent to the
statement that region I is a triangle with integer points with minimal area (=1/2).

Note that the cone generated by $\{ m_1, m_2, k \}$ centered in $(0,0,0)$
strictly includes the cone generated by the four vectors $m_i$; this is dual
to the statement that region I is included in the fan of the
conifold. The second term in (\ref{zetac}) exactly cancels the integer points that
belong
to $\{ m_1, m_2, k \}$ but not to the cone generated by the $m_i$. 

The important observation is that expansion (\ref{exp}) is valid in the region
$\{ q^{m_1}<1, q^{m_2}<1, q^{k}<1 \}$. Since the second term in (\ref{zetac})
contains the factor $q^{-k}$ before expanding it in the usual way we can
rearrange the expression (\ref{zetac}) for $Z_{\mathcal{O}}$ as
\begin{equation}
Z_{\mathcal{O}}=\frac{1}{(1-q^{m_1})(1-q^{m_2})(1-q^{k})} -
\frac{q^k}{(1-q^{m_3})(1-q^{m_4})(1-q^{k})}
\label{cancel}
\end{equation}
Now we can expand both these terms in the region: $q^{m_i}<1, \forall
i=1,\ldots 4$ and $q^{k}<1$. The factor of $q^{k}$ in front of the second term
simply translates the origin: we get the partition function that counts
integer points inside the cone $\{ m_1, m_2, k \}$ with origin $(0,0,0)$ minus
the partition function that counts integer points inside the cone $\{
m_3,m_4,k \}$ centered at the integer point $k$. Note that since $k$ is a
primitive vector, along the line with direction $k$ and passing through
$(0,0,0)$, the point $k$ is the first integer point after the origin
$(0,0,0)$. We have thus canceled all the integer points we didn't want to
count and hence $Z_{\mathcal{O}}$ is the correct partition function. A pictorial description
of equation (\ref{cancel}) is given in Figure \ref{pqweb} b); we project the
edges of the cones on the (p,q) web plane (first two coordinates). The reader
should try to imagine the process in three dimensions.

Obviously one can repeat the same process exchanging $k \leftrightarrow -k$:
we expand the second term of (\ref{zetac}) for $\{ q^{m_3}<1, q^{m_4}<1,
q^{-k}<1 \}$ and rearrange the first term of (\ref{zetac}); we see that the
same expansion for $Z_{\mathcal{O}}$ is valid in the region  $q^{m_i}<1, \forall
i=1,\ldots 4$ and $q^{-k}<1$. Combining with the previous result, we obtain
that $Z_{\mathcal{O}}$ can be expanded in the region $q^{m_i}<1, \forall
i=1,\ldots 4$, that is only the external vectors $m_i$ matter.
Obviously taking the second resolution for the conifold of Figure
\ref{conifold} one arrives at the same function $Z_{\mathcal{O}}$.

It is easy to see now how to write the partition function that counts integer
points inside a polytope obtained by moving the origins $v_1$ and $v_2$ of the
two cones at arbitrary, non coincident, points. We obtain: 
\begin{equation}
\frac{q^{v_1}}{(1-q^{m_1})(1-q^{m_2})(1-q^{k})} +
\frac{q^{v_2}}{(1-q^{m_3})(1-q^{m_4})(1-q^{-k})}
\end{equation} 
since the factors $q^{v_J}$ are simply translating the origins. For instance
in the case of $Z_{D_1}$ in Section \ref{conifoldexample} we had $v_1=(1,1,-1)$ and
$v_2=(0,0,0)$,
there called $m^0_I$ and $m^0_{II}$.
This is another explanation of formula (\ref{sumzd}).

It is not difficult to generalize this example to all cases we are interested
in. Suppose you have a rational convex polytope in $\mathbb{R}^3$ with integer
vertices; call $m^0_J$ its vertices and $m_J^{\lambda}$ the edges attached to each
vertex $m^0_J$. We normalize the $m_J^{\lambda}$ to primitive integer vectors
(all exiting from the vertex $m^0_J$). 
At each vertex suppose that the infinite cone generated by the $m_J^{\lambda}$
is of Calabi-Yau type (meaning that the dual cone has generators lying on the
plane $z=1$); as we will see later this is our case. Then one can easily compute the
partition function $Z_J(q)$ that counts integer points inside the infinite
cone with vertex in $(0,0,0)$ generated
by the vectors $m_J^{\lambda}$, with fixed $J$, for
instance by taking any resolution of the associated fan. 
Repeating the trick above, it is easy to
see that the partition function that counts integer points inside the original
polytope  
is given by the sum $\sum_J q^{m^0_J} Z_J(q)$ over all vertices $J$ of the polytope.

\begin{figure}
\begin{center}
\includegraphics[scale=0.57]{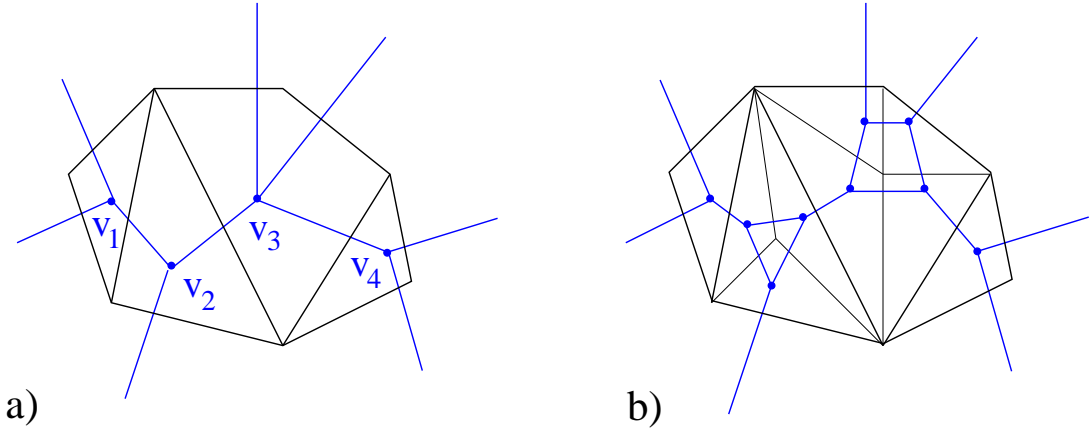} 
\caption{a) Edges of the polytope $P_D$ projected on the (p,q) web plane (in blue)
and dual resolved toric diagram (in black).  b) The same picture for $\tilde{P}_D$.}
\label{x1}
\end{center}
\end{figure}

Now we go back to the original problem of Section \ref{riemannroch}.
To fix the notation, let $c_i$ be the generic integer weights assigned to each
vertex $n_i$ of a toric diagram that define the bundle: $\mathcal{O}(\sum_i c_i
D_i)$, $i \in \Sigma(1)$; where $\Sigma(1)$ is the set of vertices $n_i=(y_i,z_i,1)$
of the toric diagram. 
Let $P_D$ the polytope in $\mathbb{R}^3$ defined by the equations:
\begin{equation}
P_D= \{ m \in \mathbb{R}^3 | \langle m , n_i \rangle + c_i \geq 0, \, \forall i \in
\Sigma(1) \}
\label{eqpd}
\end{equation}
One can compute the intersections of these planes and reconstruct the edges
and the vertices of $P_D$. There is a plane for each vertex $V_i$. An example
is reported in Figure \ref{x1} a): in this case the polytope has 7 planes and 4
vertices $v_J$.
In general $P_D$ is convex and has rational edges, however its vertices are
only rational and may not be integer. Therefore we define another convex
polytope $\tilde{P}_D$ as the convex hull of all integer points inside
$P_D$. Therefore $\tilde{P}_D \subseteq P_D$ and all integer points in $P_D$
belong also to $\tilde{P}_D$: the original problem is reduced to the problem of
counting integer points inside $\tilde{P}_D$. 

It is easy to see that $\tilde{P}_D$ can be alternatively described by adding
equations to those defining $P_D$ (\ref{eqpd}), since the infinite edges of $P_D$,
being described by rational equations, pass through integer points.  
An example is reported in Figure \ref{x1} b), where we draw projected on the (p,q)
web plane the edges of $\tilde{P}_D$ corresponding to the polytope $P_D$ in Figure
\ref{x1} a). Note that besides the 7 infinite pieces of planes delimiting $P_D$ we
have added two finite pieces of planes; in the dual description this corresponds to
refine the resolution of the toric diagram by adding two internal points
$\tilde{n}_i$, the perpendiculars to the two planes.

An important fact is that the resolution we need to describe $\tilde{P}_D$ is again
Calabi-Yau, meaning that the only planes we need to add to equations (\ref{eqpd})
are those with perpendiculars $\tilde{n}_i \in \tilde{\Sigma}(1)$ where
$\tilde{\Sigma}(1)$ is the set of integer vectors $\tilde{n}_i$ lying on the plane
of the toric diagram and inside the toric diagram. 
Consider for any integer vector $\tilde{n}_i \in \tilde{\Sigma}(1)$ the plane
$\langle m, \tilde{n}_i \rangle + a = 0$; by varying $a$ it is easy to see that for large positive $a$ the plane 
does not intersect $\tilde{P}_D$; hence there is a maximal value of $a$ for which the plane has a non empty intersection with the
closed polytope $\tilde{P}_D$. Define $\tilde{c}_i$ such value for $a$. Obviously if
$\tilde{n}_i=n_i$ coincides with an external vertex of the toric diagram we obtain
for $\tilde{c}_i$ the original value $c_i$. Note that all $\tilde{c}_i$ are integers
since $\tilde{P}_D$ has integer vertices. Now define the polytope $Q$:
\begin{equation}
Q= \{ m \in \mathbb{R}^3 | \langle m , \tilde{n}_i \rangle + \tilde{c}_i \geq 0, \,
\forall i \in \tilde{\Sigma}(1) \}
\label{qeq}
\end{equation}
with the $\tilde{c}_i$ defined as before. It is not difficult to prove that
$Q=\tilde{P}_D$. In fact from the definitions we straightforwardly obtain that:
$\tilde{P}_D \subseteq Q \subseteq P_D$. Now the convex polytope $Q$ can be seen as
the convex hull of its vertices and of the integer points along its infinite
external edges. If we prove that $Q$ has integer vertices then we would prove also
$Q \subseteq \tilde{P}_D$ since $\tilde{P}_D$ is the convex hull of all integer
points inside $P_D$; and hence $Q=\tilde{P}_D$. 

By computing the intersections of the planes in the definition (\ref{qeq}) we can
obtain the corresponding resolution of the toric diagram and the vertices of the
polytope $Q$. For example one could obtain the resolution in Figure \ref{x1} b),
where the toric diagram has been divided into 9 regions $\rho_J$, $I=1 \ldots 9$,
each corresponding to a vertex $m^0_J$ of the convex polytope $Q$. The vertex
$m^0_J$ is the intersection of the planes $\langle m, \tilde{n}_i \rangle +
\tilde{c}_i = 0$, for all the vertices $\tilde{n}_i$ of the region $\rho_J$. If the
region $\rho_J$ we are considering has no internal integer point, then by the Pick's
theorem \cite{fulton} it is a triangle with minimal area $1/2$; call its integer vertices
$\tilde{n}_1$, $\tilde{n}_2$ and $\tilde{n}_3$. Since for this triangle
$\det(\tilde{n}_1,\tilde{n}_2,\tilde{n}_3)=1$, $m^0_J$ is an integer point. Instead
if the region $\rho_J$ has integer points inside it is easy to see from the
construction of $Q$ that all the planes $\langle m, \tilde{n}_i \rangle + \tilde{c}_i = 0$, for any integer
$\tilde{n}_i$ internal to $\rho_J$, pass through the vertex $m^0_J$. Hence we can
take any complete resolution of the region $\rho_J$ into minimal triangles and
compute $m^0_J$ as the intersection of the planes associated to its three
vertices. Hence again $m^0_J$ is integer. All minimal triangles belonging to the
region $\rho_J$ identify the same $m^0_J$. 

We just proved that $Q$ has integer vertices and hence $Q=\tilde{P}_D$. Since
$\tilde{P}_D$ is Calabi-Yau, to compute the partition function $Z_D$ that counts its
integer points we can use the method derived above: 
\begin{equation}
Z_D= \sum_{\rho_J} q^{m^0_J} Z_J (q)
\end{equation}
where as before $Z_J$ is the partition function counting integer points inside the
cone with apex in $(0,0,0)$ generated by the edges $m^{\lambda}_J$ attached to
vertex $m^0_J$. 
Moreover the partition functions $Z_J$ can be computed using any complete resolution
of the regions $\rho_J$; we obtain therefore a refined resolution of the toric
diagram in minimal triangles $\sigma_I$. The resulting partition function is just
formula (\ref{sumzd}).

Note that there is some ambiguity in choosing the complete resolution of the toric
diagram into triangles $\sigma_I$; however this resolution must be compatible with
the starting $\rho_J$ resolution.
To summarize we have given an alternative proof of (\ref{sumzd}) and we have explicitly
built the integers $\tilde{c}_i$ associated with the internal points $\tilde{n}_i
\in \tilde{\Sigma}(1)$. Then the equations (\ref{qeq}) define a resolution $\rho_J$
of the toric diagram that can be further refined. Note that the two conditions given
in Section \ref{riemannroch} are naturally satisfied with this geometrical choice of
$\tilde{c}_i$; in particular convexity (\ref{convex}) follows from the fact that
$m^0_J$ are the vertices of $\tilde{P}_D$.

\end{document}